\documentclass[preprint,12pt,fleqn]{elsarticle}

\usepackage{algorithm}
\usepackage{algorithmic}
\usepackage{amssymb}
\usepackage{amsthm}
\usepackage{amsmath}
\usepackage{graphicx}
\usepackage{hyperref}

\journal{Computer Physics Communications}

\begin{document}

\begin{frontmatter}

\title{ELSI --- An Open Infrastructure for Electronic Structure Solvers}

\author[duke_mems]{Victor Wen-zhe Yu}
\author[upv]{Carmen Campos}
\author[riken]{William Dawson}
\author[icmab]{Alberto Garc\'{i}a}
\author[aalto]{Ville Havu}
\author[strathclyde]{Ben Hourahine}
\author[duke_mems]{William P. Huhn}
\author[berkeley]{Mathias Jacquelin}
\author[berkeley,uc_berkeley]{Weile Jia}
\author[argonne]{Murat Ke\c{c}eli}
\author[duke_mems]{Raul Laasner}
\author[duke_math]{Yingzhou Li}
\author[berkeley,uc_berkeley]{Lin Lin}
\author[duke_math,duke_phys,duke_chem]{Jianfeng Lu}
\author[molssi]{Jonathan Moussa}
\author[upv]{Jose E. Roman}
\author[argonne]{\'{A}lvaro V\'{a}zquez-Mayagoitia}
\author[berkeley]{Chao Yang}
\author[duke_mems,duke_chem]{Volker Blum \corref{ca}}
\ead{volker.blum@duke.edu}
\cortext[ca]{Corresponding author.}

\address[duke_mems]{Department of Mechanical Engineering and Materials Science, Duke University, Durham, NC 27708, USA}
\address[upv]{Departament de Sistemes Inform\`{a}tics i Computaci\'{o}, Universitat Polit\`{e}cnica de Val\`{e}ncia, Val\`{e}ncia, Spain}
\address[riken]{RIKEN Center for Computational Science, Kobe 650-0047, Japan}
\address[icmab]{Institut de Ci\`{e}ncia de Materials de Barcelona (ICMAB-CSIC), Bellaterra E-08193, Spain}
\address[aalto]{Department of Applied Physics, Aalto University, Aalto FI-00076, Finland}
\address[strathclyde]{SUPA, University of Strathclyde, Glasgow G4 0NG, UK}
\address[berkeley]{Computational Research Division, Lawrence Berkeley National Laboratory, Berkeley, CA 94720, USA}
\address[uc_berkeley]{Department of Mathematics, University of California, Berkeley, CA 94720, USA}
\address[argonne]{Computational Science Division, Argonne National Laboratory, Argonne, IL 60439, USA}
\address[duke_math]{Department of Mathematics, Duke University, Durham, NC 27708, USA}
\address[duke_phys]{Department of Physics, Duke University, Durham, NC 27708, USA}
\address[duke_chem]{Department of Chemistry, Duke University, Durham, NC 27708, USA}
\address[molssi]{Molecular Sciences Software Institute, Blacksburg, VA 24060, USA}

\begin{abstract}
Routine applications of electronic structure theory to molecules and periodic systems need to compute the electron density from given Hamiltonian and, in case of non-orthogonal basis sets, overlap matrices. System sizes can range from few to thousands or, in some examples, millions of atoms. Different discretization schemes (basis sets) and different system geometries (finite non-periodic vs. infinite periodic boundary conditions) yield matrices with different structures. The ELectronic Structure Infrastructure (ELSI) project provides an open-source software interface to facilitate the implementation and optimal use of high-performance solver libraries covering cubic scaling eigensolvers, linear scaling density-matrix-based algorithms, and other reduced scaling methods in between. In this paper, we present recent improvements and developments inside ELSI, mainly covering (1) new solvers connected to the interface, (2) matrix layout and communication adapted for parallel calculations of periodic and/or spin-polarized systems, (3) routines for density matrix extrapolation in geometry optimization and molecular dynamics calculations, and (4) general utilities such as parallel matrix I/O and JSON output. The ELSI interface has been integrated into four electronic structure code projects (DFTB+, DGDFT, FHI-aims, SIESTA), allowing us to rigorously benchmark the performance of the solvers on an equal footing. Based on results of a systematic set of large-scale benchmarks performed with Kohn--Sham density-functional theory and density-functional tight-binding theory, we identify factors that strongly affect the efficiency of the solvers, and propose a decision layer that assists with the solver selection process. Finally, we describe a reverse communication interface encoding matrix-free iterative solver strategies that are amenable, e.g., for use with planewave basis sets.
\end{abstract}


\end{frontmatter}

{\bf \noindent PROGRAM SUMMARY\\}
\begin{small}
\noindent
{\em Program title: ELSI Interface\\}
{\em Licensing provisions: BSD 3-clause\\}
{\em Distribution format: .tar.gz, git repository\\}
{\em Programming language: Fortran 2003, with interface to C/C++\\}
{\em External routines/libraries: BLACS, BLAS, BSEPACK (optional), EigenExa (optional), ELPA, FortJSON, LAPACK, libOMM, MPI, MAGMA (optional), MUMPS (optional), NTPoly, ParMETIS (optional), PETSc (optional), PEXSI, PT-SCOTCH (optional), ScaLAPACK, SLEPc (optional), SuperLU\_DIST\\}
{\em Operating system: Unix-like (Linux, macOS, Windows Subsystem for Linux)\\}
{\em Nature of problem: Solving the electronic structure from given Hamiltonian and overlap matrices in electronic structure calculations.\\}
{\em Solution method: ELSI provides a unified software interface to facilitate the use of various electronic structure solvers including cubic scaling dense eigensolvers, linear scaling density matrix methods, and other approaches.}
\end{small}

\section{Introduction}
\label{sec:intro}
Computer simulations based on electronic structure theory, particularly Kohn--Sham density-functional theory (KS-DFT)~\cite{dft_hohenberg_1964,dft_kohn_1965}, are facilitating scientific discoveries across a broad range of disciplines such as chemistry, physics, and materials science. In KS-DFT, the many-electron problem for the Born--Oppenheimer electronic ground state is reduced to a system of single particle equations known as the Kohn--Sham equations,
\begin{equation}
\label{eq:ks}
\hat{h}^\text{KS} \psi_l = \epsilon_l \psi_l ,
\end{equation}

\noindent where $\hat{h}^\text{KS}$ denotes the Kohn--Sham Hamiltonian, and $\psi_l$ and $\epsilon_l$ are the Kohn--Sham orbitals and their associated eigenenergies. In most computer realizations of KS-DFT, Eq.~\ref{eq:ks} is discretized by expanding $\psi_l$ with $N_\text{basis}$ basis functions $\phi_j$:
\begin{equation}
\label{eq:basis}
\psi_l(\boldsymbol{r}) = \sum_{j=1}^{N_\text{basis}} c_{jl} \phi_j(\boldsymbol{r}) ,
\end{equation}

\noindent which converts Eq.~\ref{eq:ks} into a generalized eigenproblem in a matrix form
\begin{equation}
\label{eq:evp}
\boldsymbol{H} \boldsymbol{C} = \boldsymbol{S} \boldsymbol{C} \boldsymbol{\epsilon} .
\end{equation}

\noindent Here $\boldsymbol{H}$ and $\boldsymbol{S}$ are the Hamiltonian matrix and the overlap matrix, respectively. $\boldsymbol{\epsilon}$ and $\boldsymbol{C}$ contain eigenvalues and eigenvectors of the eigensystem of $\boldsymbol{H}$ and $\boldsymbol{S}$. Solving Eq.~\ref{eq:evp} by diagonalization yields the Kohn--Sham orbitals (through Eq.~\ref{eq:basis}) and their eigenenergies, from which the electron density can be computed. Despite its success in a variety of traditional KS-DFT implementations, this approach has a computational cost that scales as O($N^3$), with $N$ being some measure of the system size. When handling complex systems consisting of many thousands of atoms, it often becomes prohibitively expensive even on today's most powerful supercomputers.

An alternative avenue to the electron density is available through the density matrix $\boldsymbol{P}$:
\begin{equation}
\label{eq:density_matrix}
p_{ij} = \sum_{l=1}^{N_\text{basis}} f_l c_{il} c_{jl}^* ,
\end{equation}

\noindent where $f_l$ is the occupation number of the $l^\text{th}$ orbital, $c_{il}$ and $c_{jl}$ are elements of the eigenvector matrix $\boldsymbol{C}$ in Eq.~\ref{eq:evp}, corresponding to coefficients of the $i^\text{th}$ and $j^\text{th}$ basis functions for the $l^\text{th}$ orbital, respectively, in Eq.~\ref{eq:basis}. The density matrix $\boldsymbol{P}$ may be computed directly from $\boldsymbol{H}$ and $\boldsymbol{S}$ without explicitly solving the eigenproblem in Eq.~\ref{eq:evp}. Since the early 1990s, density-matrix-based linear scaling algorithms~\cite{nearsightedness_kohn_1996,on_goedecker_1999,on_bowler_2012}, particularly for spatially localized basis functions, have been developed to overcome the scaling bottleneck of diagonalization. Even millions of atoms can be simulated with advanced implementations of linear scaling KS-DFT~\cite{million_bowler_2010,million_vandevondele_2012}. However, the larger computational prefactor associated with algorithms that target the density matrix without the diagonalization in Eq.~\ref{eq:evp} hinders their application in anything but very large systems. The diagonalization-based approach, in contrast, is generally applicable and highly efficient for systems comprised of up to several hundred atoms and can remain competitive up to several thousand atoms (see Sec.~\ref{sec:results}).

In the last decade, a number of new algorithms targeting the Kohn--Sham eigenproblem have emerged as software libraries, such as PEXSI (pole expansion and selected inversion)~\cite{pexsi_lin_2009,pexsi_lin_2013,pexsi_lin_2014,pexsi_jia_2017}, CheSS (Fermi operator expansion by Chebyshev polynomials)~\cite{chess_mohr_2017}, and iterative eigensolvers powered by spectrum slicing~\cite{sips_campos_2012,sips_keceli_2016,slice_li_2017,evsl_li_2019} or Chebyshev filtering~\cite{chebyshev_banerjee_2016,chebyshev_banerjee_2018,chase_winkelmann_2019}. Each of these algorithms has quite unique features, performance characteristics, and expert regimes. The crossover point between direct diagonalization and these alternative methods depends on the specifics of the simulation. It is thus a difficult task to select the optimal numerical method for a given system under study.

The ELectronic Structure Infrastructure (ELSI) project~\cite{elsi_yu_2018} provides an open-source, integrated software interface connecting electronic structure code packages to various high-performance eigensolvers and density matrix solvers. Encouragingly, the ELSI interface has attracted great interest from the community. To date, it supports five eigensolvers (ELPA~\cite{elpa_auckenthaler_2011,elpa_marek_2014,elpa_kus_2019}, SLEPc-SIPs~\cite{sips_campos_2012,sips_keceli_2016,sips_keceli_2018,slepc_hernandez_2005}, EigenExa~\cite{eigenexa_imamura_2011}, LAPACK~\cite{lapack_anderson_1999}, MAGMA~\cite{magma_tomov_2010,magma_dongarra_2014}), three density matrix solvers (libOMM~\cite{libomm_corsetti_2014}, PEXSI~\cite{pexsi_lin_2009,pexsi_lin_2013,pexsi_lin_2014,pexsi_jia_2017}, NTPoly~\cite{ntpoly_dawson_2018}), and a special purpose eigensolver targeting the Bethe--Salpeter equation (BSEPACK~\cite{bsepack_shao_2016}). ELSI has been integrated with four electronic structure packages (DFTB+~\cite{dftb_hourahine_2020}, DGDFT~\cite{dgdft_hu_2015}, FHI-aims~\cite{fhiaims_blum_2009}, SIESTA~\cite{siesta_soler_2002}). It is open for adoption and modification by any other electronic structure code. In addition, ELSI is included in the Electronic Structure Library (ESL) project~\cite{esl_oliveira_2020}, a distribution of shared open-source libraries in the electronic structure community. In this paper, we present an update of the ELSI software from its 1.0.0 release (May 2017)~\cite{elsi_yu_2018} to the current 2.5.0 release (February 2020), covering newly added solvers, support of periodic and spin-polarized calculations, density matrix extrapolation routines for geometry optimization and molecular dynamics (MD) calculations, parallel matrix I/O, and a native Fortran library for JSON output. The ELSI interface and its integration into existing electronic structure codes enable us to rigorously benchmark the performance of multiple solvers on an equal footing. We report a systematic set of large-scale benchmarks performed with Kohn--Sham density-functional theory and density-functional tight-binding theory. Factors that strongly affect the efficiency of the solvers are identified and analyzed, based on which we propose a ``decision layer'' that assists users in selecting an appropriate solver for an arbitrary problem. Finally, we outline a reverse communication interface (RCI) encoding matrix-free iterative solver strategies, currently including the Davidson method~\cite{davidson_davidson_1975,davidson_sleijpen_1996}, the orbital minimization method (OMM)~\cite{libomm_corsetti_2014,omm_lu_2017}, the projected preconditioned conjugate gradient (PPCG) method~\cite{ppcg_vecharynski_2015}, and the Chebyshev filtering method~\cite{chebyshev_banerjee_2016,chebyshev_zhou_2006}. This feature would benefit especially planewave-based DFT, but also other applications where only a few eigenvectors of a high-dimensional matrix are needed.

\section{Upgraded and New Features in ELSI (from 2017 to 2019)}
\label{sec:feature}
In this section, we first briefly review the basic idea behind the ELSI software interface and its design, fundamentals of which were described in Ref.~\cite{elsi_yu_2018}. We then elaborate on new features added to ELSI since our previous publication, including support for new solvers and new matrix formats, API extension for managing multiple eigenproblems simultaneously across a given number of MPI tasks, new functions needed for routine electronic structure simulations such as the extrapolation of normalized density matrices between different underlying system geometries, and general utilities such as parallel matrix I/O and a native Fortran library for JSON output. An RCI framework for iterative eigensolvers will be discussed in Sec.~\ref{subsec:rci} and will be published in more detail separately. A new CMake-based build system for ELSI will additionally be discussed in \ref{app:cmake}.

\subsection{Review of ELSI API}
\label{subsec:review}
The ELSI infrastructure is intended for the rapid integration of a variety of eigensolvers and density matrix solvers into existing electronic structure codes. The API of ELSI is designed to be compatible with the workflow of an electronic structure code. There are three key steps to use ELSI in a code implementing the self-consistent field (SCF) method: (1) At the beginning of an SCF cycle, the electronic structure code initializes ELSI by calling the subroutine \textbf{elsi\_init}, which returns an ``ELSI handle'' storing all runtime parameters of the ELSI interface and the solvers. ``ELSI handle'' is a derived data type defined in Fortran. C/C++ code can also access it via the iso\_c\_binding module of Fortran 2003. (2) Within the SCF cycle, the electronic structure code uses one of the ELSI solver interfaces to solve the Kohn--Sham eigenproblem by an eigensolver, or to compute the density matrix by a density matrix solver. This is done by calling \textbf{elsi\_\{ev$\vert$dm\}\_\{real$\vert$complex\}\{\_sparse\}} for real or complex, dense or sparse matrices. (3) After the SCF cycle converges, the electronic structure code finalizes ELSI by calling \textbf{elsi\_finalize}. At any point after the initialization of an ELSI handle and before its finalization, the electronic structure code can tune the parameters of the interface and the solvers. A detailed explanation of the ELSI workflow is available in Ref.~\cite{elsi_yu_2018} (see Fig. 3 and Algorithm 1 of that reference). Algorithms~\ref{alg:elsi_old} and \ref{alg:elsi_new} in Sec.~\ref{subsec:reinit} also give a brief overview.

\subsection{Solver Updates}
\label{subsec:solver}
As of its 2.5.0 release, ELSI supports shared-memory eigensolvers LAPACK~\cite{lapack_anderson_1999} and MAGMA~\cite{magma_tomov_2010,magma_dongarra_2014}, distributed-memory eigensolvers ELPA~\cite{elpa_auckenthaler_2011,elpa_marek_2014,elpa_kus_2019}, SLEPc-SIPs~\cite{sips_campos_2012,sips_keceli_2016,sips_keceli_2018,slepc_hernandez_2005}, and EigenExa~\cite{eigenexa_imamura_2011}, distributed-memory density matrix solvers PEXSI~\cite{pexsi_lin_2009,pexsi_lin_2013,pexsi_lin_2014,pexsi_jia_2017}, libOMM~\cite{libomm_corsetti_2014}, and NTPoly~\cite{ntpoly_dawson_2018}, and the distributed-memory Bethe--Salpeter equation solver BSEPACK~\cite{bsepack_shao_2016}. In the following sections, we cover a few algorithmic and technical aspects of the upgraded PEXSI solver and the newly added NTPoly and SLEPc-SIPs solvers. The reader is referred to the publications cited above for more details of the solvers supported in ELSI.

\subsubsection{PEXSI (upgraded)}
\label{subsubsec:pexsi}
The pole expansion and selected inversion (PEXSI) algorithm~\cite{pexsi_lin_2009,pexsi_lin_2013} belongs to the category of Fermi operator expansion (FOE) methods. PEXSI expands the density matrix $\boldsymbol{P}$ as a sum of rational matrix functions:
\begin{equation}
\label{eq:pexsi}
\boldsymbol{P} = \sum_l \mathbb{I}\text{m} \left( \frac{\omega_l}{\boldsymbol{H} - (z_l + \mu) \boldsymbol{S}} \right) ,
\end{equation}

\noindent where $\mu$ is the chemical potential of the system, $\{z_l\}$ and $\{\omega_l\}$ are complex shifts (``poles'') and weights of the expansion terms. Here we assume matrices $\boldsymbol{H}$ and $\boldsymbol{S}$ to be real symmetric for simplicity. The formulation becomes slightly but not fundamentally different when $\boldsymbol{H}$ and $\boldsymbol{S}$ are complex Hermitian matrices.

For spatially localized basis functions and KS-DFT, only a subset of the elements of $\boldsymbol{H}$ and $\boldsymbol{S}$ (the set for which two basis functions overlap) will be non-zero and thus only a subset of the elements of the object $(\boldsymbol{H} - (z_l + \mu) \boldsymbol{S})^{-1}$ in Eq.~\ref{eq:pexsi}, i.e. those corresponding to non-zero elements of $\boldsymbol{H}$ and $\boldsymbol{S}$, need to be computed for PEXSI. This is done using the parallel selected inversion method~\cite{pselinv_jacquelin_2016,pselinv_jacquelin_2018}. The computational complexity of Eq.~\ref{eq:pexsi} depends on the dimensionality of the system: O($N$), O($N^{1.5}$), and O($N^2$) for 1D, 2D, and 3D systems, respectively, with $N$ being the size of the system. This favorable scaling relies on the sparsity of the matrices which may be achieved with localized basis functions, but does not rely on the existence of an energy gap.

With PEXSI version 0.10, which was used in ELSI 1.0.0~\cite{elsi_yu_2018}, 40 to 100 poles are sufficient for the result obtained from PEXSI to be fully comparable (within $10^{-5}$ eV/atom~\cite{elsi_yu_2018}) to that obtained from diagonalization. The chemical potential is obtained by a Newton type method that may need several iterations to converge at the beginning of an SCF cycle. Eq.~\ref{eq:pexsi} is evaluated once in each of the iterations. As the SCF cycle proceeds and the electron density stabilizes, the quasi-Newton search can usually converge in one iteration.

The recently released PEXSI version 1.2 incorporates significant improvements over previous versions. First, the minimax rational approximation of the Fermi--Dirac distribution~\cite{pole_moussa_2016} becomes the default pole expansion method, reducing the number of terms needed in Eq.~\ref{eq:pexsi} to around 10 $\sim$ 30. Second, the Newton type method for finding the chemical potential is replaced by an algorithm, summarized below, that simultaneously improves the efficiency and robustness~\cite{pexsi_jia_2017}. Instead of computing the exact chemical potential $\mu$ in every SCF iteration, the new algorithm tracks the upper and lower bounds of $\mu$, denoted as $\mu_\text{max}$ and $\mu_\text{min}$. Then, an interpolation is carried out to estimate $\mu$ from $\mu_\text{max}$ and $\mu_\text{min}$. As the SCF cycle converges, $\mu_\text{max}$ and $\mu_\text{min}$ are guaranteed to converge from both sides towards the final, exact $\mu$. Owing to the reduced number of poles and the elimination of the Newton iterations, PEXSI 1.2 shows a significant speed-up over previous versions.

\subsubsection{NTPoly (new)}
\label{subsubsec:ntpoly}
Density matrix purification is an established way to achieve linear scaling in electronic structure theory~\cite{on_goedecker_1999,on_bowler_2012}. Assuming an orthogonal basis set, the density matrix $\boldsymbol{P}$ at zero temperature is known to satisfy three conditions:
\begin{equation}
\label{eq:density_matrix_conditions}
\begin{split}
& \text{Hermitian}: \boldsymbol{P} = \boldsymbol{P}^* ,\\
& \text{Normalized}: \text{Tr}(\boldsymbol{P}) = N_\text{electron} ,\\
& \text{Idempotent}: \boldsymbol{P} = \boldsymbol{P}^2 .
\end{split}
\end{equation}

\noindent Since an eigenproblem in a non-orthogonal basis can be transformed to an orthogonal basis by a decomposition of the overlap matrix $\boldsymbol{S}$, e.g. the L\"{o}wdin decomposition:
\begin{equation}
\label{eq:ntpoly_lowdin}
\begin{split}
\boldsymbol{\tilde{H}} \boldsymbol{C} & = \boldsymbol{C} \boldsymbol{\epsilon} ,\\
\boldsymbol{\tilde{H}} & = \boldsymbol{S}^{-1/2} \boldsymbol{H} \boldsymbol{S}^{-1/2} ,
\end{split}
\end{equation}

\noindent we will stick to the assumption of an orthogonal basis set throughout this subsection.

Starting from an initial guess of the density matrix $\boldsymbol{P}_0$, density matrix purification methods iteratively update the density matrix by applying
\begin{equation}
\label{eq:ntpoly_purification}
\boldsymbol{P}_\text{n+1} = f(\boldsymbol{P}_\text{n}) ,
\end{equation}

\noindent where $\boldsymbol{P}_\text{n}$ is the density matrix in the n$^\text{th}$ purification iteration, $\boldsymbol{P}_\text{n+1}$ is the density matrix in the (n+1)$^\text{th}$ iteration, and $f(\boldsymbol{P})$ is usually a matrix polynomial, chosen to guarantee that $\boldsymbol{P}_\text{n}$ rapidly and stably converges to a matrix satisfying Eq.~\ref{eq:density_matrix_conditions}.

For a matrix to satisfy the idempotent condition in Eq.~\ref{eq:density_matrix_conditions}, its eigenvalues can only be 0 and 1. The initial guess $\boldsymbol{P}_0$ that enters Eq.~\ref{eq:ntpoly_purification} is usually obtained by scaling the Hamiltonian matrix to make its eigenvalues lie in between 0 and 1~\cite{purification_niklasson_2002}:
\begin{equation}
\label{eq:ntpoly_scale_hamiltonian}
\boldsymbol{P}_0 = \frac{\epsilon_\text{max} \boldsymbol{I} - \boldsymbol{H}}{\epsilon_\text{max} - \epsilon_\text{min}} ,
\end{equation}

\noindent where $\epsilon_\text{max}$ and $\epsilon_\text{min}$ are the (estimated) maximum and minimum eigenvalues of $\boldsymbol{H}$. Then, a number of choices for $f(\boldsymbol{P})$ are able to drive all eigenvalues towards 0 or 1. We refer the reader to Refs.~\cite{on_goedecker_1999,on_bowler_2012} and references therein for a review of density matrix purification algorithms.

For systems with a significant energy gap, the magnitude of the density matrix elements $p_{ij}$ in Eq.~\ref{eq:density_matrix} decreases exponentially with respect to the distance between the $i^\text{th}$ and $j^\text{th}$ basis functions, $\vert \boldsymbol{r}_i - \boldsymbol{r}_j \vert$~\cite{nearsightedness_kohn_1996}. A cutoff distance $r_\text{cutoff}$ is often used with linear scaling methods to truncate the density matrix, such that when $\vert \boldsymbol{r}_i - \boldsymbol{r}_j \vert > r_\text{cutoff}$, the corresponding matrix element $p_{ij}$ is forced to be zero. This truncation leads to a highly sparse density matrix for large systems. Alternatively, the sparsity of the density matrix may be enforced by dynamically filtering out matrix elements smaller than a predefined threshold. Given sufficiently large systems with a proper gap, the density matrix will be highly sparse, and the computational complexity of density matrix purification is O($N$).

Exploiting its sparse, massively parallel matrix-matrix multiplication kernel, the NTPoly library~\cite{ntpoly_dawson_2018} implements the canonical purification algorithm~\cite{purification_palser_1998} that preserves the number of electrons throughout the purification iterations, the trace resetting purification methods~\cite{purification_niklasson_2002} that successively refine the number of electrons to the target value, and the generalized canonical purification method~\cite{purification_truflandier_2016} that relies on the relationship between the electron density matrix and the hole density matrix to accelerate convergence. A comparison of these methods is given in Ref.~\cite{ntpoly_dawson_2018}. In NTPoly, the sparsity of $\{\boldsymbol{P}_\text{n}\}$ is ensured by setting any matrix element smaller than a predefined threshold to zero. This approach allows for an easier integration with electronic structure codes, as it does not require any knowledge on the physical system or the basis functions.

\subsubsection{SLEPc-SIPs (new)}
\label{subsubsec:sips}
The shift-and-invert spectral transformation method, implemented in the SLEPc library~\cite{sips_campos_2012,slepc_hernandez_2005}, transforms the eigenproblem in Eq.~\ref{eq:evp} by shifting the eigenspectrum:
\begin{equation}
\label{eq:shift}
(\boldsymbol{H} - \sigma \boldsymbol{S}) \boldsymbol{C} = \boldsymbol{S} \boldsymbol{C} (\boldsymbol{\epsilon} - \sigma \boldsymbol{I}) ,
\end{equation}

\noindent where $\sigma$ is a shift and $\boldsymbol{I}$ is the identity matrix. This shifted eigenproblem is converted to the standard form by inverting ($\boldsymbol{H} - \sigma \boldsymbol{S})$ and $(\boldsymbol{\epsilon} - \sigma \boldsymbol{I})$:
\begin{equation}
\label{eq:invert}
(\boldsymbol{H} - \sigma \boldsymbol{S})^{-1} \boldsymbol{S} \boldsymbol{C} = \boldsymbol{C} (\boldsymbol{\epsilon} - \sigma \boldsymbol{I})^{-1} ,
\end{equation}

\noindent which is a standard eigenproblem
\begin{equation}
\label{eq:sips_standard_eigenproblem}
\begin{split}
\boldsymbol{\tilde{H}} \boldsymbol{C} & = \boldsymbol{C} \boldsymbol{\tilde{\epsilon}} ,\\
\boldsymbol{\tilde{H}} & = (\boldsymbol{H} - \sigma \boldsymbol{S})^{-1} \boldsymbol{S} ,\\
\boldsymbol{\tilde{\epsilon}} & = (\boldsymbol{\epsilon} - \sigma \boldsymbol{I})^{-1} .
\end{split}
\end{equation}

\noindent The transformed standard eigenproblem in Eq.~\ref{eq:sips_standard_eigenproblem} is solved by an iterative Krylov--Schur algorithm in SLEPc~\cite{sips_campos_2012}. If a shift can be chosen to be close to the target eigenvalue, the shift-and-invert transformation augments the magnitude of the eigenvalue, accelerating the solution of the standard eigenproblem in Eq.~\ref{eq:sips_standard_eigenproblem}.

The shift-and-invert transformation becomes less effective when many eigenvalues need to be computed, because not all of them are close to the shift. The spectrum slicing technique is an advanced modification that employs multiple shifts to compute all eigenvalues contained in an interval. A large interval can be partitioned into independent slices and solved in parallel. The parallel spectrum slicing method implemented in SLEPc-SIPs~\cite{sips_campos_2012,sips_keceli_2016} partitions the eigenspectrum of an eigenproblem into $N_\text{slice}$ slices. Correspondingly, the computer processes involved in the calculation are split into $N_\text{slice}$ groups, so that each slice is handled by one group. Thanks to this additional layer of parallelism, this approach has the potential to exhibit enhanced scalability over the direct diagonalization method, which has been demonstrated in calculations on the density-functional-based tight-binding (DFTB) level~\cite{sips_keceli_2016} as well as on the KS-DFT level~\cite{sips_keceli_2018}. Additionally, SLEPc-SIPs should greatly outperform direct diagonalization methods in cases where only a small fraction of the eigenspectrum is wanted.

\subsection{Matrix Format Updates}
\label{subsec:matrix}
In software implementations of electronic structure theory, matrices can be stored in any form. ELSI, as a unified solver interface, must be able to handle matrices stored in commonly used formats, and must be able to convert between these formats. Two matrix formats were implemented in the previous version of ELSI, namely the 2D block-cyclic distributed dense format (BLACS\_DENSE) and the 1D block distributed compressed sparse column (CSC) sparse format (PEXSI\_CSC)~\cite{elsi_yu_2018}. They have enabled the use of the ELPA, libOMM, and PEXSI solvers in the electronic structure codes FHI-aims and DGDFT. In the present version of ELSI, two additional matrix formats are available, namely the 1D block-cyclic distributed CSC sparse format (SIESTA\_CSC) and the generic coordinate sparse format (GENERIC\_COO). The 1D and 2D block-cyclic distribution schemes BLACS\_DENSE and SIESTA\_CSC essentially cover all variants of block/cyclic/block-cyclic distributions, as pure cyclic and pure block distributions can be viewed as special cases of the general block-cyclic distribution. The GENERIC\_COO format is designed to support any data distribution scheme that might be employed to store matrices in an electronic structure code. On each process, a list of triplets is constructed, containing local matrix elements and their global row and column indices. Based on the indices, ELSI then redistributes the matrix to the format needed by the chosen solver. The triplet list can be arbitrarily distributed, sorted or unsorted.

Conversions between any two of the supported matrix formats are implemented with MPI. In order to avoid unnecessary data communication, all zeros in a matrix are always ignored in the redistribution process. As reported in Ref.~\cite{elsi_yu_2018}, time spent on the conversion is negligible relative to the actual computation time.

\subsection{Interface Extension for Spin-Polarized and Periodic Systems}
\label{subsec:spin_kpt}
The parallelization strategy behind the ELSI interface depends on the physical system being simulated. The base case is an isolated system in vacuum, e.g. free atoms, molecules, clusters, without spin-polarization. In this case, there is one eigenproblem (Eq.~\ref{eq:evp}) in each iteration of an SCF cycle. The strategy to tackle this single eigenproblem has been detailed in Ref.~\cite{elsi_yu_2018}. We here introduce how spin-polarized and periodic systems are supported in ELSI.

When a spin-polarized periodic system is considered, Eq.~\ref{eq:evp} will have an index $\alpha$ denoting the spin channel, and an index \textbf{\textit{k}} denoting points in reciprocal space:
\begin{equation}
\label{eq:spin_kpt_eigenproblem}
\boldsymbol{H}_{\boldsymbol{k}}^\alpha \boldsymbol{C}_{\boldsymbol{k}}^\alpha = \boldsymbol{S}_{\boldsymbol{k}} \boldsymbol{C}_{\boldsymbol{k}}^\alpha \boldsymbol{\epsilon}_{\boldsymbol{k}}^\alpha .
\end{equation}

\noindent Because of the periodicity, it is sufficient to study \textbf{\textit{k}} within a single primitive unit cell in reciprocal space, usually the first Brillouin zone (1BZ). The physical quantities are represented by integrals in 1BZ. Take the electron density $n (\boldsymbol{r})$ as an example:
\begin{equation}
\label{eq:brillouin_zone}
n (\boldsymbol{r}) = \sum_{\alpha=1}^{N_\text{spin}} \sum_{l=1}^{N_\text{basis}} \int_\text{1BZ} f_{l \boldsymbol{k}}^\alpha \psi_{l \boldsymbol{k}}^{\alpha*} (\boldsymbol{r}) \psi_{l \boldsymbol{k}}^\alpha (\boldsymbol{r}) d^3 \boldsymbol{k} ,
\end{equation}

\noindent which is approximated by using a finite mesh of \textbf{\textit{k}}-points in 1BZ:
\begin{equation}
\label{eq:k_grid}
n (\boldsymbol{r}) \approx \sum_{\alpha=1}^{N_\text{spin}} \sum_{n=1}^{N_\text{kpt}} w_{\boldsymbol{k}_n} \sum_{l=1}^{N_\text{basis}} f_{l \boldsymbol{k}_n}^\alpha \psi_{l \boldsymbol{k}_n}^{\alpha*} (\boldsymbol{r}) \psi_{l \boldsymbol{k}_n}^\alpha (\boldsymbol{r}) .
\end{equation}

\noindent Here, $\psi_{l \boldsymbol{k}_n}^\alpha$ and $f_{l \boldsymbol{k}_n}^\alpha$ are the l$^\text{th}$ band at the n$^\text{th}$ \textbf{\textit{k}}-point in the $\alpha$ spin channel and its occupation number, $w_{\boldsymbol{k}_n}$ is the weight of the n$^\text{th}$ \textbf{\textit{k}}-point, $N_\text{spin}$, $N_\text{kpt}$, and $N_\text{basis}$ are the number of spin channels, \textbf{\textit{k}}-points, and basis functions, respectively. The weights of all \textbf{\textit{k}}-points add up to 1. A denser grid of \textbf{\textit{k}}-points leads to a more accurate description of 1BZ, at the price of higher computational cost. If the unit cell in real space is small, an accurate description of the electronic structure requires thousands of \textbf{\textit{k}}-points or even more. If the unit cell in real space is large, the Brillouin zone may already be well-represented by the origin of the reciprocal space, known as the $\Gamma$ point. The Hamiltonian and overlap matrices for multiple \textbf{\textit{k}}-points have a block-diagonal structure. Each block on the diagonal corresponds to an eigenproblem of one \textbf{\textit{k}}-point.

\begin{figure*}[ht!]
\centering
\includegraphics[width=0.6\textwidth]{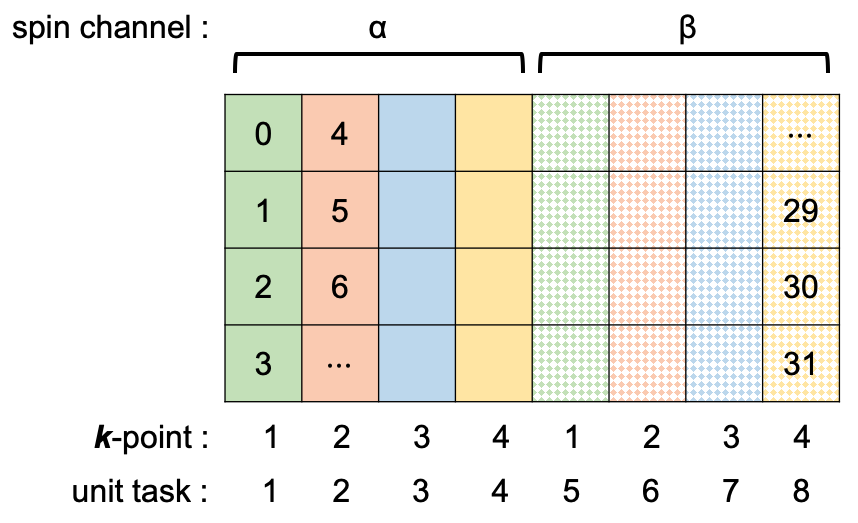}
\caption{Parallel calculation of spin-polarized and periodic systems in ELSI exemplified with a fictitious system of two spin channels and four \textbf{\textit{k}}-points, solved by 32 parallel processes. Each square box represents one process, with its index (0, 1, ..., 31) labelled inside. The 32 processes are divided into eight process groups, with four processes in each. Spin channel $\alpha$ and $\beta$ are indicated by boxes with solid and chessboard background, respectively. \textbf{\textit{k}}-points 1, 2, 3, and 4 are indicated by green, red, blue, and yellow boxes, respectively.}
\label{fig:spin}
\end{figure*}

In total, there are $N_\text{kpt} \times N_\text{spin}$ eigenproblems in Eq.~\ref{eq:spin_kpt_eigenproblem}. They can be solved in an embarrassingly parallel fashion. In ELSI, eigenproblems in Eq.~\ref{eq:spin_kpt_eigenproblem} are considered as equivalent ``unit tasks'', which are set up by the electronic structure code. The available computer processes are divided into $N_\text{kpt} \times N_\text{spin}$ groups, each of which is responsible for one unit task. An example with two spin channels ($\alpha$ and $\beta$) and four \textbf{\textit{k}}-points (1, 2, 3, 4) solved by 32 parallel processes (0, 1, ..., 31) is given in Fig.~\ref{fig:spin}. There are eight unit tasks in this example. Each unit task is handled by four processes in a ``solve'' phase followed by a ``reduction'' phase, which are further explained below. In an actual calculation, the total number of processes is not restricted to be a multiple of the number of eigenproblems $N_\text{kpt} \times N_\text{spin}$. Process groups can have different numbers of processes, although a uniform partition usually leads to the optimal load balance.

Computational steps carried out in the ``solve'' phase include:

\begin{itemize}
\item ELPA, SLEPc-SIPs, and EigenExa: Solve the eigenproblems in Eq.~\ref{eq:spin_kpt_eigenproblem}.

\item PEXSI: Perform the pole expansion for each unit task at a chemical potential uniform across all tasks. Compute the number of electrons for each unit task.

\item libOMM and NTPoly: Perform orbital minimization or density matrix purification for each unit task to obtain density matrices.
\end{itemize}

Computational steps carried out in the ``reduction'' phase include:

\begin{itemize}
\item ELPA, SLEPc-SIPs, and EigenExa: Eigensolutions for the unit tasks are coupled by the normalization condition of the number of electrons:
\begin{equation}
\label{eq:normalization}
N_\text{electron} = \sum_{n=1}^{N_\text{kpt}} \sum_{\alpha=1}^{N_\text{spin}} \sum_{l=1}^{N_\text{basis}} w_{\boldsymbol{k}_n} f_{l \boldsymbol{k}_n}^\alpha ,
\end{equation}

\noindent The eigenvalues solved for by each unit task are collected across all the tasks, for the determination of the chemical potential and occupation numbers.

\item PEXSI: The total number of electrons is computed as a weighted summation over the number of electrons solved for all unit tasks.

\item libOMM and NTPoly: None.
\end{itemize}

To use the ELSI interface in a spin-polarized and/or periodic calculation, two MPI communicators should be passed into ELSI, for data communication within a unit task in the ``solve'' phase and communication between all unit tasks in the ``reduction'' phase, respectively. We note that the parallelization strategy described in this subsection can only be applied when there are more processes than unit tasks. When the number of processes is small, the ELSI interface can be set up in such a way that each process invokes the eigensolver in LAPACK or MAGMA to solve a few unit tasks sequentially. This is referred to as the ``SINGLE\_PROC'' parallelization mode in Ref.~\cite{elsi_yu_2018}.

\subsection{Calculation of the Energy-Weighted Density Matrix}
\label{subsec:edm}
In the context of geometry optimization or molecular dynamics (collectively referred to as ``geometry calculations'' hereafter), forces (derivatives of the total energy) are needed to evaluate the movement of atoms. There is a so-called ``Pulay'' force term that originates as localized basis functions move with the atoms~\cite{pulay_pulay_1969}. One ingredient needed to compute the ``Pulay'' force is the energy-weighted density matrix $\boldsymbol{Q}$:
\begin{equation}
\label{eq:energy_density_matrix}
q_{ij} = \sum_{l=1}^{N_\text{basis}} \epsilon_l f_l c_{il} c_{jl}^* ,
\end{equation}

\noindent which is the density matrix $\boldsymbol{P}$ in Eq.~\ref{eq:density_matrix} weighted by eigenvalues of the orbitals. Density matrix solvers compute $\boldsymbol{Q}$ directly from $\boldsymbol{H}$, $\boldsymbol{S}$, and $\boldsymbol{P}$. In ELSI, $\boldsymbol{Q}$ is not computed at the time when a density matrix solver is invoked through \textbf{elsi\_dm\_\{real$\vert$complex\}\{\_sparse\}}. Since forces are typically only needed near or after the end of an SCF cycle, always computing $\boldsymbol{Q}$ together with $\boldsymbol{P}$ is unnecessary. Instead, \textbf{elsi\_get\_edm\_\{real$\vert$complex\}\{\_sparse\}} is dedicated to retrieving $\boldsymbol{Q}$ whenever it is needed.

\subsection{Calculation of the Electronic Entropy}
\label{subsec:entropy}
At zero temperature, the occupation number $\{f_{l \boldsymbol{k}}^\alpha\}$ of an orbital in a metallic system drops abruptly from 1 (omitting spin degeneracy) to 0 at the Fermi energy. This is undesired in two aspects. First, to accurately integrate this discontinuous function in, e.g, Eq.~\ref{eq:k_grid}, a very fine grid of \textbf{\textit{k}}-points is required to sample the Brillouin zone. More \textbf{\textit{k}}-points lead to a more accurate description of the Brillouin zone, at the price of proportionally increasing computational cost. Second, when an orbital suddenly changes from unoccupied to occupied, or vice versa, its contribution to the electron density (Eq.~\ref{eq:k_grid}) suddenly appears or disappears, leading to instabilities in the convergence of an SCF cycle.

In practical calculations, a metallic system is usually treated at a higher electronic temperature by means of ``smearing'' of the density of states, which makes the occupation function decrease smoothly from 1 to 0 around the Fermi level. This continuous function can be integrated more accurately with a relatively coarse grid of \textbf{\textit{k}}-points. In addition, using a smearing function helps stabilize the SCF convergence for any system with a zero or small energy gap~\cite{vasp_kresse_1996,smearing_rabuck_1999}. A side effect is that the energy functional being minimized no longer corresponds to the total energy $E_\text{tot}$ at zero temperature, but the free energy $E_\text{free} = E_\text{tot} - TS$, where $T$ and $S$ are the electronic temperature and the electronic entropy, respectively~\cite{entropy_gillan_1989,entropy_weinert_1992}. Correspondingly, ``forces'' needed for geometry calculations should be computed as gradients of the free energy, instead of the regular total energy. In fact, in the analytical gradient of the free energy, the term involving the derivative of the fractional occupation numbers with respect to the atomic displacement exactly cancels with the corresponding term originating from the derivative of the electronic entropy~\cite{entropy_weinert_1992}.

In ELSI, we have implemented the Fermi--Dirac~\cite{smearing_mermin_1965}, Gaussian~\cite{smearing_fu_1983}, Methfessel--Paxton~\cite{smearing_methfessel_1989}, and Marzari--Vanderbilt~\cite{smearing_marzari_1999} smearing functions. Given the exact number of electrons and the eigenenergies of the orbitals (computed by the ELPA, SLEPc-SIPs, EigenExa, LAPACK, or MAGMA eigensolver), the occupation numbers, chemical potential, and electronic entropy can be calculated with any one of the four smearing functions. The electronic entropy term is currently unavailable from ELSI when using a density matrix solver. The PEXSI solver is capable of directly computing the free energy density matrix~\cite{pexsi_lin_2013}, from which the free energy and Fermi--Dirac entropy can be deduced.

\subsection{Reinitialization of ELSI}
\label{subsec:reinit}
When the atomic positions get updated in geometry optimization or molecular dynamics calculations, localized basis functions move together with atoms, leading to a new overlap matrix $\boldsymbol{S}_1$, and a sparsity pattern different from that of the previous overlap matrix $\boldsymbol{S}_0$. In the previous version of ELSI, the user was responsible for finalizing ELSI and reinitializing it for an updated geometry, as shown in Algorithm~\ref{alg:elsi_old}. This guarantees that outdated information is not carried over to the updated geometry. However, this also discards information that can be reused, such as the MPI setup and most of the solver-specific settings. To maximize the reuse of information between geometry steps, and to minimize the coding effort on the electronic structure code side, we have introduced the \textbf{elsi\_reinit} subroutine to reinitialize an instance of ELSI. The usage of \textbf{elsi\_reinit} in geometry calculations is demonstrated in Algorithm~\ref{alg:elsi_new}. Compared to \textbf{ELSI\_OLD}, the initialization and finalization of ELSI are moved out of the geometry loop, avoiding the repeated re-creation of ELSI instances. A new geometry step is indicated by calling \textbf{elsi\_reinit}, which instructs ELSI to flush geometry-related variables and arrays that cannot be safely reused in the new geometry step, mainly the overlap matrix and its sparsity pattern. Other information is kept within the ELSI instance and reused throughout multiple geometry steps.

\begin{algorithm}[ht!]
\caption{Usage of the previous version of the ELSI interface~\cite{elsi_yu_2018} in geometry optimization and molecular dynamics calculations. For simplicity, tasks belonging to the electronic structure code, e.g., the construction of electron density and the integration of Hamiltonian, are not shown.}
\begin{algorithmic}
\STATE{\textbf{procedure ELSI\_OLD}}
    \WHILE{(geometry not converged)}
        \STATE{call elsi\_init}
        \STATE{call elsi\_set\_*}
        \WHILE{(SCF not converged)}
            \STATE{call elsi\_\{ev$\vert$dm\}}
        \ENDWHILE
        \STATE{call elsi\_finalize}
    \ENDWHILE
\end{algorithmic}
\label{alg:elsi_old}
\end{algorithm}

\begin{algorithm}[ht!]
\caption{Usage of the present version of the ELSI interface in geometry optimization and molecular dynamics calculations. Compared to the \textbf{ELSI\_OLD} procedure in \ref{alg:elsi_old}, repeated re-creation of ELSI instances is avoided by using \textbf{elsi\_reinit}.}
\begin{algorithmic}
\STATE{\textbf{procedure ELSI\_NEW}}
    \STATE{call elsi\_init}
    \STATE{call elsi\_set\_*}
    \WHILE{(geometry not converged)}
        \WHILE{(SCF not converged)}
            \STATE{call elsi\_\{ev$\vert$dm\}}
        \ENDWHILE
        \STATE{call elsi\_reinit}
    \ENDWHILE
    \STATE{call elsi\_finalize}
\end{algorithmic}
\label{alg:elsi_new}
\end{algorithm}

\subsection{Extrapolation of Wavefunctions and the Density Matrix}
\label{subsec:extrapolation}
In a single point total energy calculation, a simple way to construct an initial guess for the electron density is to use a superposition of free atom densities. In geometry calculations, the initial guess in the (n+1)$^\text{th}$ geometry step can be improved by reusing the wavefunctions or density matrix calculated in the n$^\text{th}$ geometry step. However, due to the movement of atoms and localized basis functions around them, wavefunctions obtained in the n$^\text{th}$ geometry step are no longer orthonormalized in the (n+1)$^\text{th}$ geometry step. Similarly, the density matrix from the n$^\text{th}$ geometry step $\boldsymbol{P}_0$ is no longer normalized with respect to the new overlap matrix $\boldsymbol{S}_1$, i.e., $\text{Tr}(\boldsymbol{P}_0 \boldsymbol{S}_1) \neq N_\text{electron}$ and $\boldsymbol{P}_0 \neq \boldsymbol{P}_0 \boldsymbol{S}_1 \boldsymbol{P}_0$.

The wavefunction coefficients, i.e., eigenvectors in Eq.~\ref{eq:evp}, can be reorthonormalized with respect to $\boldsymbol{S}_1$. This is implemented in ELSI with the Gram--Schmidt algorithm. For the density matrix, we decompose the previous and current overlap matrices $\boldsymbol{S}_0$ and $\boldsymbol{S}_1$, then extrapolate the previous density matrix $\boldsymbol{P}_0$ to a new density matrix $\boldsymbol{P}_1$~\cite{lowdin_lowdin_1950,extrapolation_mezey_1997,extrapolation_niklasson_2010}. With a Cholesky decomposition of the overlap matrices, the process is:
\begin{equation}
\label{eq:cholesky_extrapolation}
\begin{split}
\boldsymbol{S}_0 & = \boldsymbol{L}_0 \boldsymbol{L}_0^*, \\
\boldsymbol{S}_1 & = \boldsymbol{L}_1 \boldsymbol{L}_1^*, \\
\boldsymbol{P}_1 & = (\boldsymbol{L}_1^{-1})^* \boldsymbol{L}_0^* \boldsymbol{P}_0 \boldsymbol{L}_0 \boldsymbol{L}_1^{-1} ,
\end{split}
\end{equation}

\noindent An equivalent extrapolation formula exists when using the L\"{o}wdin decomposition. The common idea is transforming $\boldsymbol{P}_0$ to an orthogonal basis, then to the new non-orthogonal basis. Both algorithms are implemented for real and complex, dense and sparse matrices.

\subsection{Parallel Matrix I/O}
\label{subsec:io}
Matrices from a given electronic structure calculation are often reusable in further computational steps -- e.g., in the simplest case, to restart a particular calculation that could not be completed within a single run, or that needed to be revisited for future data extraction tasks. In such cases, it is helpful to write some information (e.g. the current atomic structure and density matrix) to file, which serves as a checkpoint to restart a calculation. When ELSI runs in parallel with multiple MPI tasks, matrices are distributed across tasks. The idea of writing a distributed matrix into $N_\text{MPI}$ separate files, where $N_\text{MPI}$ is the number of MPI tasks, is not promising due to the difficulty of post-processing a large number of files. Therefore, parallel matrix I/O (input/output) routines are preferred. Other use cases of this functionality include, e.g., sharing matrices between collaborating teams and testing the solvers with pre-generated matrices.

In the past, we have encountered some difficulties in using existing parallel I/O libraries~\cite{hdf5_folk_1999,netcdf4_li_2003} with thousands of MPI tasks on various supercomputers. This might be connected to overhead introduced by complex hierarchical data structures implemented in high level I/O libraries. Therefore, the data structure in ELSI is simply arrays that represent matrices, if necessary, accompanied by a few integers to define the dimensions of the matrices. It is more straightforward to directly employ the parallel I/O functionality defined in the MPI standard~\cite{mpiio_corbett_1996}. Our implementation of matrix I/O in ELSI is built around MPI\_File\_\{read$\vert$write\}\_at\_all, which allows distributed data to be written to (read from) a single file, as if the data was gathered onto a single MPI task then written to one file (read from one file by one MPI task then scattered to all). The optimal I/O performance, both with MPI I/O and in general, is obtained by making large and contiguous requests to access the file system, rather than small, non-contiguous, or random requests. Therefore, before writing a matrix to file, we redistribute it to a 1D block distribution. This guarantees that each task writes a contiguous chunk of data to a contiguous piece of file. Similarly, a matrix read from file is in a 1D block distribution. Whenever needed, it can be redistributed automatically to one of the supported distributions.

\subsection{JSON Output via the FortJSON Library}
\label{subsec:json}
The workflow of benchmarking multiple solvers across different electronic structure codes is greatly accelerated by using a consistent output format that is easily processed by external scripting utilities. To aid in accelerating this workflow when using ELSI, we provide JSON output from ELSI using FortJSON, an open source JSON library written by ELSI developers for Fortran 2003 code bases. All benchmarks presented in Sec.~\ref{sec:results} were output using FortJSON.

FortJSON is bundled as part of the ELSI package and is available to codes linked against ELSI. Alternatively, it may be downloaded as a standalone package from the ELSI GitLab server and installed using a CMake build system. It uses a handle-based structure similar to ELSI's, in which the user initializes a FortJSON handle to open a JSON file, calls a consistent API to write to the JSON file, and finalizes the handle to finish output. JSON files written by FortJSON are fully compliant with the ``ECMA-404 The JSON Data Interchange'' standard~\cite{json_2017}. FortJSON is, in principle, a standalone library that can be used to facilitate JSON output in any other context as well.

\subsection{Summary of New Features in ELSI}
\label{subsec:update}
As a summary of this section, all major changes of ELSI since Ref.~\cite{elsi_yu_2018} are listed below.

\begin{itemize}
\item PEXSI v1.2: Default number of poles reduced to 20 without sacrificing accuracy; efficient and robust strategy for finding the chemical potential. (Sec.~\ref{subsubsec:pexsi})
\item NTPoly: Linear scaling density matrix purification methods (canonical purification, 2$^\text{nd}$ and 4$^\text{th}$ order trace resetting purification, generalized hole-particle canonical purification) in the NTPoly library. (Sec.~\ref{subsubsec:ntpoly})
\item SLEPc-SIPs: Shift-and-invert parallel spectrum slicing eigensolver in the SLEPc library. (Sec.~\ref{subsubsec:sips})
\item EigenExa: Tridiagonalization and penta-diagonalization eigensolvers in the EigenExa library.
\item MAGMA: GPU-accelerated one-stage and two-stage tridiagonalization eigensolvers in the MAGMA library.
\item BSEPACK: Distributed-memory eigensolver in the BSEPACK library, specifically targeting the Bethe--Salpeter equation.
\item 1D block-cyclic distributed compressed sparse column matrix format (SIESTA\_CSC). (Sec.~\ref{subsec:matrix})
\item Arbitrarily distributed coordinate sparse matrix format (GENERIC\_COO). (Sec.~\ref{subsec:matrix})
\item Interface extension for spin-polarized and periodic systems. (Sec.~\ref{subsec:spin_kpt})
\item Energy-weighted density matrix available for all supported solvers and matrix formats. (Sec.~\ref{subsec:edm})
\item Electronic entropy with Fermi--Dirac, Gaussian, Methfessel--Paxton, and Marzari--Vanderbilt broadening schemes. (Sec.~\ref{subsec:entropy})
\item Reinitialization of ELSI between geometry steps. (Sec.~\ref{subsec:reinit})
\item Density matrix extrapolation between geometry steps. (Sec.~\ref{subsec:extrapolation})
\item Gram--Schmidt orthogonalization of eigenvectors between geometry steps. (Sec.~\ref{subsec:extrapolation})
\item MPI I/O based parallel matrix I/O for dense and sparse matrices. (Sec.~\ref{subsec:io})
\item JSON output via the FortJSON library. (Sec.~\ref{subsec:json})
\item Iterative eigensolvers in a reverse communication interface (RCI) framework. (Sec.~\ref{subsec:rci})
\item CMake build system. (\ref{app:cmake})
\end{itemize}

\section{Benchmarks and Discussions}
\label{sec:results}
Since the previous version, we have successfully integrated the ELSI interface into two additional electronic structure projects, DFTB+ and SIESTA. It is now possible for the users of DFTB+, DGDFT, FHI-aims, and SIESTA to easily switch between eigensolvers and density matrix solvers supported in ELSI by specifying a single keyword in their package-specific input files. Depending on the needs and knowledge level of the users, ELSI can be used as a black box solution, or can be fine-tuned to allow for optimal performance in certain circumstances. Developments, enhancements, and fixes made in ELSI and the solvers are immediately available for users of these electronic structure codes. Knowledge gained from ELSI usage in one code often benefits users of other codes.

The ELSI interface and its integration with electronic structure codes enable us to rigorously benchmark the performance of different solvers on an equal footing. We present here a large set of cross-solver, cross-code benchmarks as an essential step to settle the respective efficiency of different solvers in different regimes. Testing Hamiltonian and overlap matrices are constructed from actual electronic structure calculations, at the levels of all-electron full-potential KS-DFT (with the FHI-aims code~\cite{fhiaims_blum_2009}), pseudopotential KS-DFT (with the SIESTA code~\cite{siesta_soler_2002}), and DFTB (with the DFTB+ code~\cite{dftb_hourahine_2020}).

We select eight benchmark atomic structure models, which include small and large models ranging from a hundred to tens of thousands of atoms, 1D, 2D, and 3D materials, light and heavy elements, and gapped and gapless systems. According to their composition and dimensionality, the structures are classified into three sets: the carbon set (Sec.~\ref{subsec:carbon}), the heavy set (Sec.~\ref{subsec:heavy}), and the bulk set (Sec.~\ref{subsec:bulk}). Our comparison of solver performance focuses on key computational steps of the solvers that are repeated in every SCF iteration. These repeated steps are: for ELPA, transforming the generalized eigenproblem in Eq.~\ref{eq:evp} to the standard form, solving the standard eigenproblem, back-transforming the eigenvectors, and constructing the density matrix via Eq.~\ref{eq:density_matrix}; for PEXSI, computing the density matrix via Eq.~\ref{eq:pexsi}; for NTPoly, transforming the generalized eigenproblem in Eq.~\ref{eq:evp} to the standard form by Eq.~\ref{eq:ntpoly_lowdin}, computing the density matrix via Eq.~\ref{eq:ntpoly_purification}, and back-transforming the density matrix. There are other computationally expensive steps that are needed only once for a fixed geometry, such as the Cholesky or LU factorization of the overlap matrix. They have less significant effects on the total time of an SCF cycle, thus are not discussed here.

As explained in Sec.~\ref{subsec:spin_kpt}, there are multiple eigenproblems in a spin-polarized system and/or a periodic system with multiple \textbf{\textit{k}}-points. Since these eigenproblems can be solved fully in parallel, the total computational cost can be roughly predicted from the cost of one eigenproblem. Therefore, we stick to a spin-non-polarized, periodic setting with a $1 \times 1 \times 1$ \textbf{\textit{k}}-grid (i.e., $\Gamma$ point only) in all our benchmarks. Table~\ref{tab:settings} summarizes solver-specific parameters used in our benchmarks. For ELPA, we use the two-stage diagonalization algorithm~\cite{elpa_marek_2014,2step_bischof_1994}. A block size of 16 is chosen for the block-cyclic matrix distribution, which usually leads to the optimal performance of the two-stage ELPA solver~\cite{elpa_marek_2014,elpa_cook_2018}. For PEXSI, we use 20 poles ($z_l$ in Eq.~\ref{eq:pexsi}), derived from the minimax rational approximation~\cite{pole_moussa_2016}. For NTPoly, we use the $4^\text{th}$ order trace-resetting purification method~\cite{purification_niklasson_2002} with the truncation parameter (matrix elements below which are discarded in order to maintain the sparsity of the matrices) being $10^{-5}$ and the convergence criterion $10^{-2}$. With these settings, the maximum difference in total energy for a given geometry is 0.5 $\mu$eV/atom between results obtained with ELPA and PEXSI, and 32.4 $\mu$eV/atom between results obtained with ELPA and NTPoly, indicating excellent accuracy of the solvers. The rest of this section will mainly focus on the performance of the solvers. Complete input files used in this section are available at the ELSI GitLab server~\cite{input_2019}.

\begin{table*}[ht!]
\centering
\caption{List of solver-specific parameters, including the diagonalization algorithm and block size for ELPA, the pole expansion algorithm and number of poles for PEXSI, the density matrix purification algorithm, truncation parameter, and convergence criterion for NTPoly. The same settings are employed across all benchmarks reported in this paper.}
\footnotesize
\begin{tabular}{l c c}
\hline
\hline
solver & parameter & value \\
\hline
ELPA   & algorithm       & two-stage diagonalization \\
ELPA   & block size      & 16 \\
\hline
PEXSI  & algorithm       & minimax rational approximation \\
PEXSI  & number of poles & 20 \\
\hline
NTPoly & algorithm       & $4^\text{th}$ order trace-resetting \\
NTPoly & truncation      & $10^{-5}$ \\
NTPoly & convergence     & $10^{-2}$ \\
\hline
\hline
\end{tabular}
\label{tab:settings}
\end{table*}

\subsection{Benchmark Set I: Carbon Allotropes}
\label{subsec:carbon}
\begin{figure*}[ht!]
\centering
\includegraphics[width=0.6\textwidth]{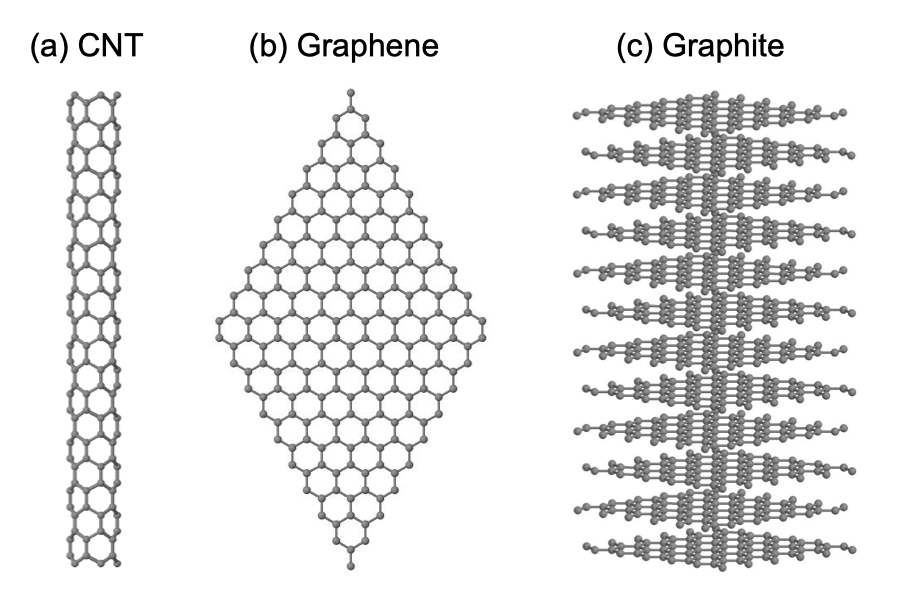}
\caption{Atomic structures of (a) 1D carbon nanotube (CNT), (b) 2D graphene, and (c) 3D graphite.}
\label{fig:carbon}
\end{figure*}

In the carbon benchmark set, we construct models of three carbon allotropes, namely (quasi) 1D carbon nanotube, 2D graphene, and 3D graphite, as shown in Fig.~\ref{fig:carbon} (a) (b) and (c), respectively. We run KS-DFT calculations with the PBE~\cite{pbe_perdew_1996} semi-local exchange-correlation functional with two software packages, namely FHI-aims and SIESTA. Input geometries are made identical in FHI-aims and SIESTA calculations. Predefined ``tier 1'' and ``DZP'' (double-zeta plus polarization) basis sets are used in FHI-aims and SIESTA, respectively. Both basis sets contain the minimal sp functions and an additional set of spd functions. The number of basis functions per atom in SIESTA is one fewer than that in FHI-aims, because of the use of pseudopotentials in SIESTA. The dimensions of the models, the number of basis functions, and the sparsity factor of the corresponding matrices are summarized in Table~\ref{tab:carbon}.

\begin{table*}[ht!]
\centering
\caption{Dimension of structures in the carbon benchmark set, and sparsity of the Hamiltonian matrix in the calculation. $\tilde{n}_\text{basis}$ is the average number of basis functions per atom. $N_\text{atom}$ is the number of atoms. The global dimension of the Hamiltonian, overlap etc. matrices is equal to the total number of basis functions $N_\text{basis} = \tilde{n}_\text{basis} \times N_\text{atom}$. Sparsity is defined as $N_\text{zero}/N_\text{basis}^2$, with $N_\text{zero}$ being the number of zero matrix elements. A minimal + spd basis set is used to describe carbon atoms in FHI-aims and SIESTA.}
\footnotesize
\begin{tabular}{c c c c c c c}
\hline
\hline
code & system & $\tilde{n}_\text{basis}$ & $N_\text{atom}$ & sparsity (\%) \\
\hline
FHI-aims & CNT      & 14 &   800 --  6,400 & 91.3 -- 98.9 \\
FHI-aims & graphene & 14 &   800 --  7,200 & 90.5 -- 99.4 \\
FHI-aims & graphite & 14 &   864 --  6,912 & 75.1 -- 96.6 \\
\hline
SIESTA   & CNT      & 13 &   800 --  6,400 & 95.0 -- 99.4 \\
SIESTA   & graphene & 13 &   800 --  7,200 & 94.5 -- 99.5 \\
SIESTA   & graphite & 13 &   864 --  6,912 & 92.5 -- 99.1 \\
\hline
\hline
\end{tabular}
\label{tab:carbon}
\end{table*}

Calculations of the carbon set were performed on the Cray XC30 supercomputer Edison at National Energy Research Scientific Computing Center (NERSC). Each node of Edison was equipped with two 12-core Intel Ivy Bridge processors. 80 nodes were fully exploited by launching 24 MPI tasks on each node, yielding 1,920 MPI tasks in total. No OpenMP parallelization was employed.

\begin{figure*}[ht!]
\centering
\includegraphics[width=0.99\textwidth]{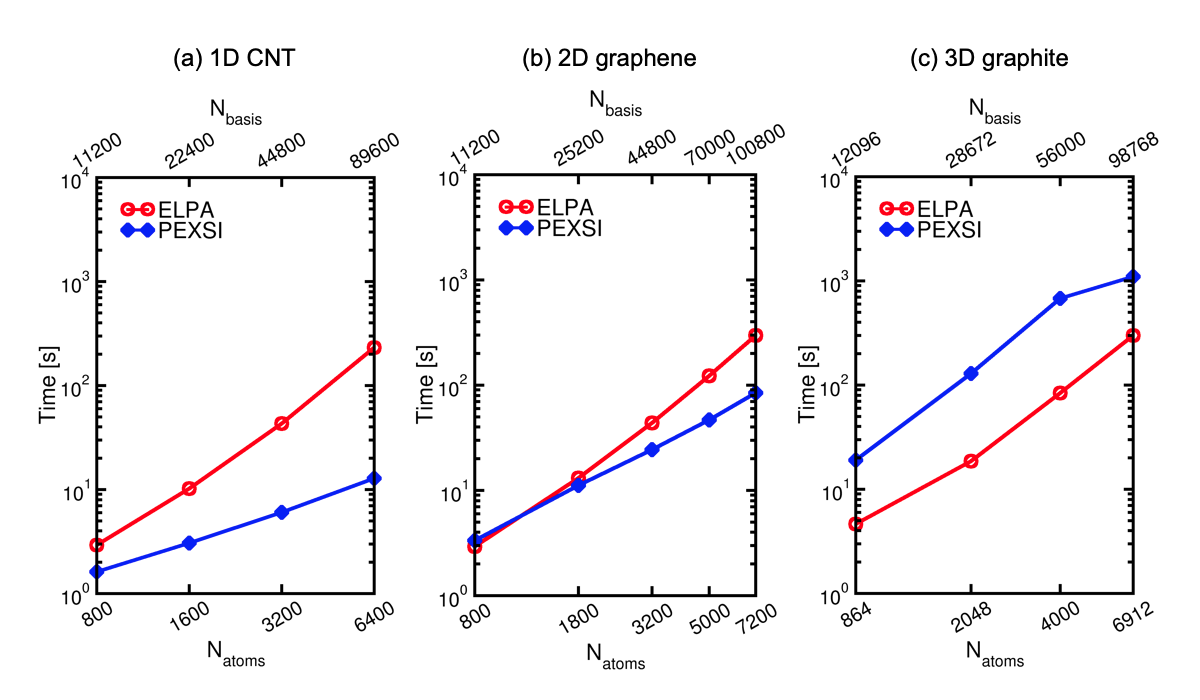}
\caption{Performance of key steps in ELPA and PEXSI for (a) carbon nanotube models, (b) graphene models, (c) graphite models, as a function of the number of atoms. For ELPA, the key steps include transforming the eigenproblem from the generalized form to the standard form, solving the standard eigenproblem, back-transforming the eigenvectors to the generalized form, and constructing the density matrix; for PEXSI, evaluating the pole expansion and selected inversion in Eq.~\ref{eq:pexsi} and assembling the density matrix. ELPA computes all eigenvalues and 42.9\% of the eigenvectors. Results in this figure are obtained by running the FHI-aims code on the Edison supercomputer with 1,920 CPU cores (MPI tasks).}
\label{fig:carbon_f}
\end{figure*}

In Fig.~\ref{fig:carbon_f}, we compare the performance of the ELPA and PEXSI solvers in all-electron KS-DFT calculations with the FHI-aims code. Fig.~\ref{fig:carbon_f} (a) shows the wallclock time needed for ELPA to solve for 42.9\% of the eigenspectrum, versus the time needed for PEXSI to compute the density matrix, in calculations of 1D carbon nanotube models consisting of 800 to 6,400 atoms (11,200 to 89,600 basis functions). PEXSI consistently outperforms ELPA in these calculations. The benefit of using PEXSI becomes more significant as the size of the system increases, owing to the O($N$) computational complexity of PEXSI for 1D systems. Fig.~\ref{fig:carbon_f} (b) shows the same comparison between ELPA and PEXSI for 2D graphene models consisting of 800 to 7,200 atoms (11,200 to 100,800 basis functions). These examples are identical to those reported in Fig. 7 (a) of Ref.~\cite{elsi_yu_2018}, except that newer versions of ELPA and PEXSI are used here. In both Fig.~\ref{fig:carbon_f} (b) and Ref.~\cite{elsi_yu_2018}, there is a crossover point between ELPA and PEXSI. The improvements of PEXSI discussed in Sec.~\ref{subsubsec:pexsi}, especially the reduction in the number of poles~\cite{pexsi_jia_2017,pole_moussa_2016}, contribute to a speed-up of PEXSI by a factor of two, bringing down the crossover point from 3,000 atoms to about 1,000 atoms. Again, it is increasingly beneficial to use PEXSI beyond 1,000 atoms, which scales as O($N^{1.5}$) for 2D systems. The same comparison for 3D graphite models consisting of 864 to 6,912 atoms (12,096 to 96,768 basis functions) is shown in Fig.~\ref{fig:carbon_f} (c), where ELPA outperforms PEXSI. The performance of ELPA with fixed computational resources should only depend on the size of the eigenproblem, hence the roughly constant time to solution of ELPA for problems of similar size in Fig.~\ref{fig:carbon_f} (a) (b) and (c). In contrast, the performance of PEXSI heavily depends on the dimensionality of the system, favoring low dimensional systems. This can be attributed to the high sparsity of Hamiltonian matrices and generalized Cholesky factors in such systems. As quantified in the last column of Table~\ref{tab:carbon}, given the same number of atoms and basis functions, the Hamiltonian matrix is more sparse in lower dimensional systems where the surface-to-volume ratio is higher. Basis functions belonging to atoms on the surface have less overlap with the other functions, yielding more zero elements in the overlap and Hamiltonian matrices.

\begin{figure*}[ht!]
\centering
\includegraphics[width=0.99\textwidth]{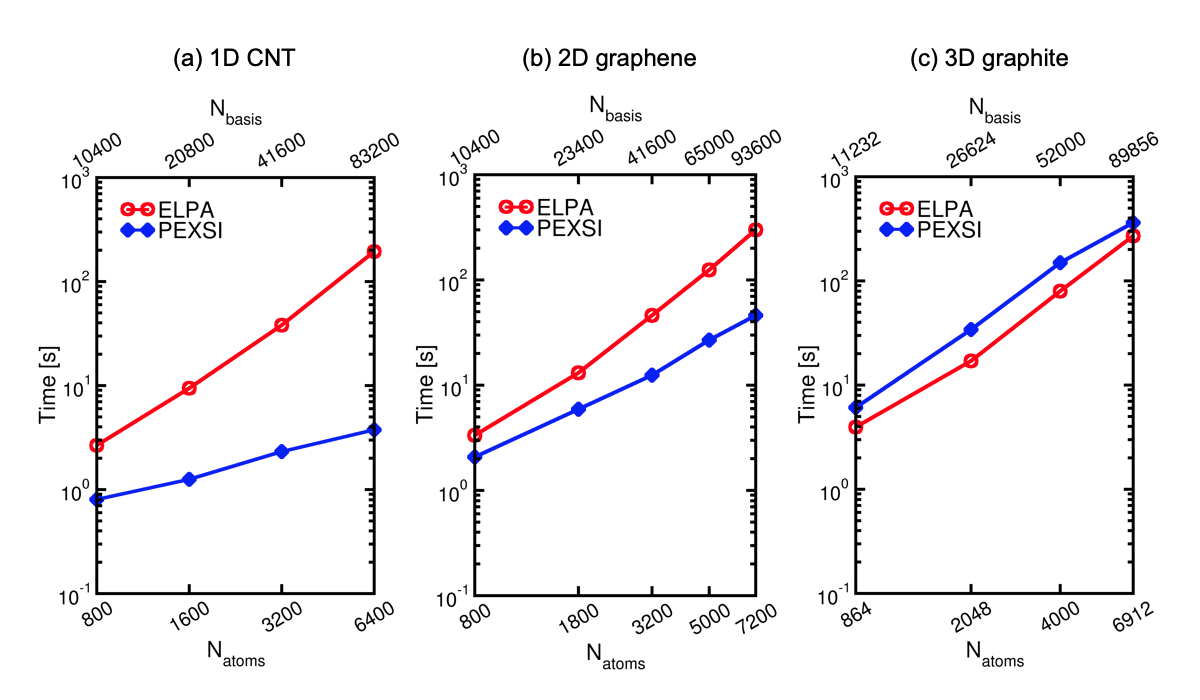}
\caption{Performance of key steps in ELPA and PEXSI for (a) carbon nanotube models, (b) graphene models, (c) graphite models, as a function of the number of atoms. ELPA computes all eigenvalues and 50\% of the eigenvectors. Results in this figure are obtained by running the SIESTA code on the Edison supercomputer with 1,920 CPU cores (MPI tasks).}
\label{fig:carbon_s}
\end{figure*}

The same performance characteristics of ELPA and PEXSI are reproduced in pseudopotential KS-DFT calculations with the SIESTA code. Basis sets, although constructed and tuned differently, are of similar size (SIESTA has one fewer basis function per atom because the 1s state is treated by a pseudopotential). The most notable difference is the width of the eigenspectrum of $\boldsymbol{H}$ and $\boldsymbol{S}$. Timings measured from calculations of 1D carbon nanotube, 2D graphene, and 3D graphite models are shown in Fig.~\ref{fig:carbon_s} (a) (b) and (c), respectively. Again, PEXSI is computationally more efficient than ELPA in 1D and 2D calculations, but not in 3D calculations. For low dimensional systems, PEXSI consistently outperforms ELPA due to the reduced asymptotic complexity, and is particularly promising for very large systems.

Owing to the all-electron formalism implemented in FHI-aims and the pseudopotential formalism implemented in SIESTA, the lowest energy level in FHI-aims, -275.99 eV, is lower than the lowest level in SIESTA, -24.34 eV, by an order of magnitude. By comparing Fig.~\ref{fig:carbon_f} with Fig.~\ref{fig:carbon_s}, we confirm that the performance of PEXSI is only mildly affected by the eigenspectrum width. In general, the runtime of PEXSI in Fig.~\ref{fig:carbon_s} is lower than that in Fig.~\ref{fig:carbon_f} for any given system. This can be explained by the fact that SIESTA matrices are slightly smaller and more sparse than FHI-aims matrices, as detailed in Table~\ref{tab:carbon}. For instance, for the graphite model with 864 atoms, the sparsity factor of the $\boldsymbol{H}$ matrix in SIESTA is 92.5\%, i.e. 7.5\% matrix elements are non-zero. The $\boldsymbol{H}$ matrix in FHI-aims contains 24.9\% non-zero elements, which makes PEXSI slower in FHI-aims calculations.

\subsection{Benchmark Set II: Heavy Elements}
\label{subsec:heavy}
\begin{figure*}[ht!]
\centering
\includegraphics[width=0.6\textwidth]{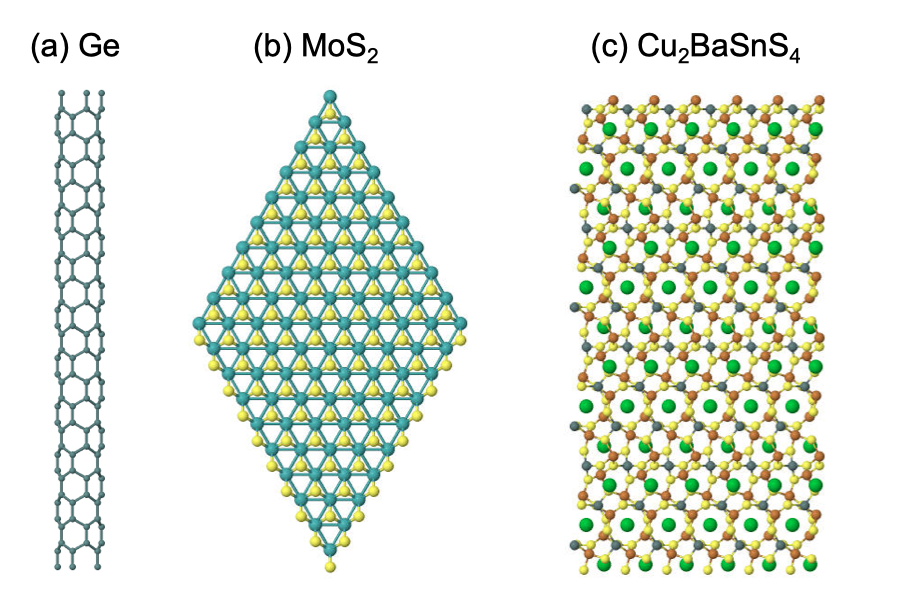}
\caption{Atomic structures of (a) 1D Ge nanotube, (b) 2D MoS$_2$ monolayer, and (c) 3D Cu$_2$BaSnS$_4$.}
\label{fig:heavy}
\end{figure*}

The heavy benchmark set is comprised of (quasi) 1D germanium nanotubes, 2D MoS$_2$ monolayers, and 3D Cu$_2$BaSnS$_4$ supercells, as shown in Fig.~\ref{fig:heavy} (a) (b) and (c), respectively. Species in this set are intentionally chosen to be heavier than carbon (atomic number: S 16, Cu 29, Ge 32, Mo 42, Sn 50, Ba 56). KS-DFT calculations with the PBE functional are carried out using the all-electron full-potential FHI-aims code, yielding eigenproblems with wide eigenspectra. For instance, the lowest eigenvalue in Cu$_2$BaSnS$_4$ calculations is -38877.57 eV, whereas the lowest value in carbon calculations is only -275.99 eV. For all species, the predefined ``tier 1'' basis sets~\cite{fhiaims_blum_2009} are employed, yielding approximately twice as many as basis functions per atom compared to the carbon set. To keep the total number of basis functions comparable to that in the carbon set, models in the heavy set contain fewer atoms. Approximately, the size of matrices in the heavy set ranges from a few thousand to over one hundred thousand, close to the size of matrices in the carbon set. The dimensions of the models, the number of basis functions, and the sparsity factor of the corresponding matrices are summarized in Table~\ref{tab:heavy}.

\begin{table*}[ht!]
\centering
\caption{Dimension of structures in the heavy benchmark set, and sparsity of the Hamiltonian matrices in the calculations. See caption of Table~\ref{tab:carbon} for definitions of $\tilde{n}_\text{basis}$, $N_\text{atom}$, and sparsity. Minimal + spd basis sets are used to describe germanium and tin atoms. Minimal + spdf basis sets are used to describe sulfur, copper, molybdenum, and barium atoms. These basis sets are predefined as ``tier 1'' in FHI-aims. Please refer to Ref.~\cite{fhiaims_blum_2009} for details on the rationale for basis set definition in FHI-aims.}
\footnotesize
\begin{tabular}{c c c c c c c}
\hline
\hline
code & system & $\tilde{n}_\text{basis}$ & $N_\text{atom}$ & sparsity (\%) \\
\hline
FHI-aims &       Ge nanotube &   27 & 400 -- 3,200 & 93.6 -- 99.2 \\
FHI-aims & MoS$_2$ monolayer & 25.3 & 300 -- 4,800 & 83.9 -- 99.4 \\
FHI-aims &   Cu$_2$BaSnS$_4$ & 26.8 & 192 -- 3,000 & 67.8 -- 97.7 \\
\hline
\hline
\end{tabular}
\label{tab:heavy}
\end{table*}

Calculations of the heavy set were performed on the Cray XC40 supercomputer Cori (Haswell partition) at NERSC. Each node of Cori is equipped with two 16-core Intel Haswell processors. 80 nodes were fully exploited by launching 32 MPI tasks on each node, yielding 2,560 MPI tasks in total. No OpenMP parallelization was employed.

\begin{figure*}[ht!]
\centering
\includegraphics[width=0.99\textwidth]{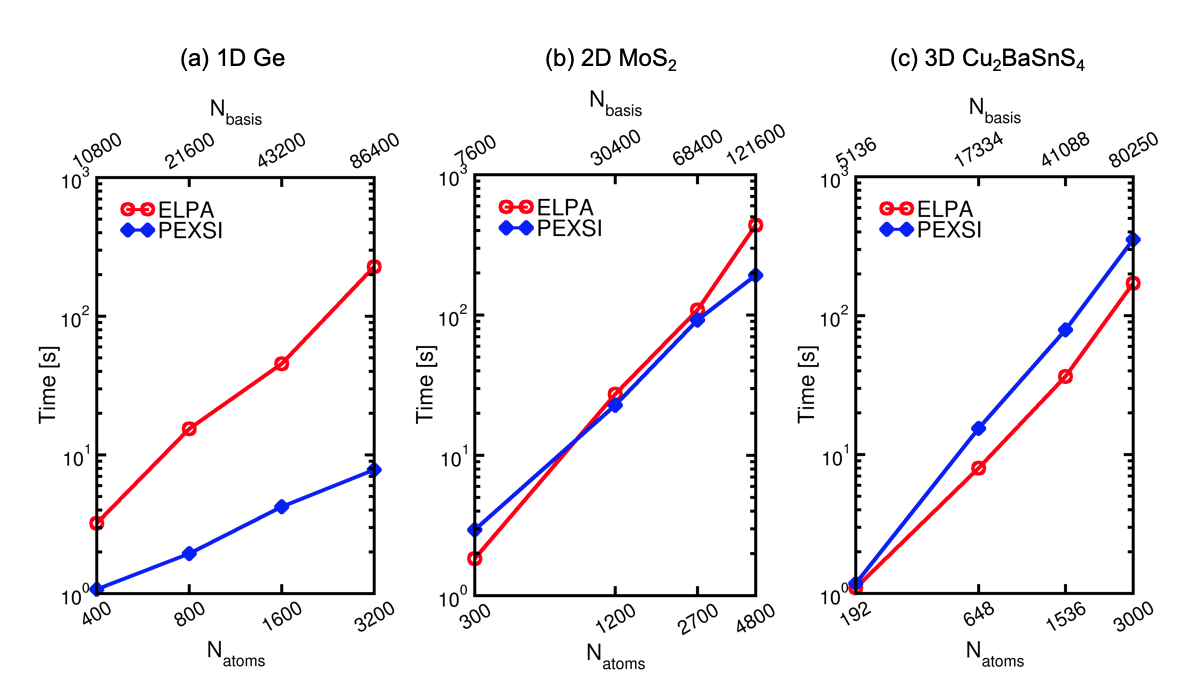}
\caption{Performance of key steps in ELPA and PEXSI for (a) Ge nanotube models, (b) MoS$_2$ monolayer models, (c) Cu$_2$BaSnS$_4$ models, as a function of the number of atoms. ELPA computes all the eigenvalues, and 79.6\%, 63.8\%, and 69.2\% of the eigenvectors for the Ge nanotube models, MoS$_2$ monolayer models, and Cu$_2$BaSnS$_4$ models, respectively. Results in this figure are obtained by running the FHI-aims code on the Cori supercomputer (Haswell partition) with 2,560 CPU cores (MPI tasks).}
\label{fig:heavy_f}
\end{figure*}

Fig.~\ref{fig:heavy_f} (a) (b) and (c) show the performance of the ELPA and PEXSI solvers in FHI-aims all-electron calculations of 1D Ge nanotube models (400 to 3,200 atoms; 10,800 to 86,400 basis functions), 2D MoS$_2$ models (300 to 4,800 atoms; 7,600 to 121,600 basis functions), and 3D Cu$_2$BaSnS$_4$ models (192 to 3,000 atoms; 5,136 to 80,250 basis functions), respectively. The wider eigenspectra and more basis functions per atom in the heavy set do not appear to have a significant impact on the relative performance of the ELPA and PEXSI solvers. The dimensionality and sparsity of the system, as in the carbon set, can greatly influence the efficiency of PEXSI. In particular, PEXSI outperforms ELPA in all 1D Ge nanotube calculations and in 2D MoS$_2$ calculations with 1,200 and more atoms. ELPA is still faster for small and dense systems, including 2D MoS$_2$ calculations with fewer than a thousand atoms and all 3D Cu$_2$BaSnS$_4$ calculations.

\subsection{Benchmark Set III: Bulk Materials}
\label{subsec:bulk}
\begin{figure*}[ht!]
\centering
\includegraphics[width=0.66\textwidth]{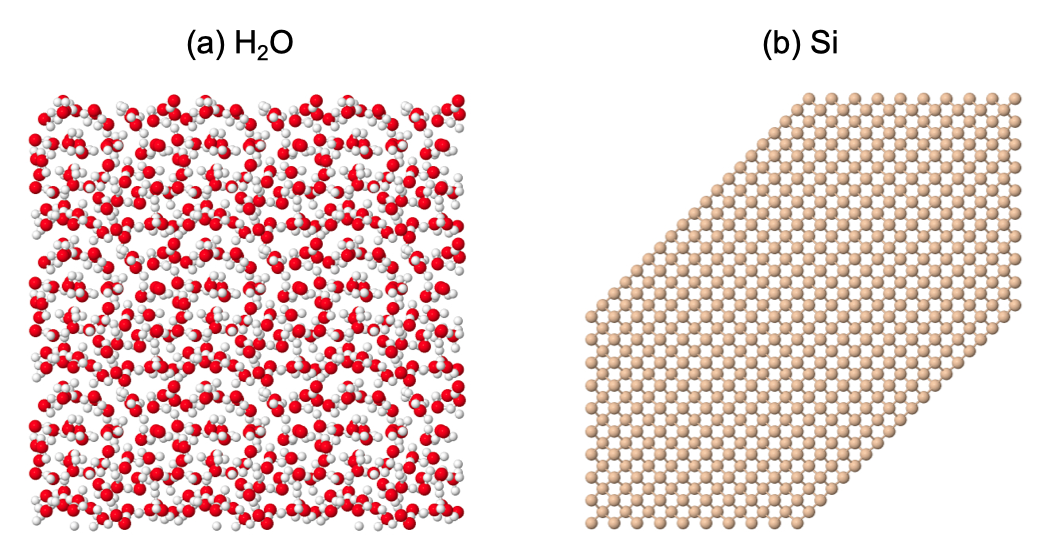}
\caption{Atomic structures of (a) water and (b) silicon.}
\label{fig:bulk}
\end{figure*}

Finally, we explore the possibility to accelerate large, 3D calculations using linear scaling density matrix purification methods implemented in the NTPoly library. A series of non-self-consistent-charge DFTB calculations of water clusters and silicon supercells, shown in Fig.~\ref{fig:bulk}, are carried out with the DFTB+ code. For hydrogen, oxygen, and silicon atoms, minimal s, sp, and sp basis sets are employed, respectively. The dimensions of the models, the number of basis functions, and the sparsity factor of the corresponding matrices are summarized in Table~\ref{tab:bulk}.

\begin{table*}[ht!]
\centering
\caption{Dimension of structures in the bulk benchmark set, and sparsity of the Hamiltonian matrices in the calculations. See caption of Table~\ref{tab:carbon} for definitions of $\tilde{n}_\text{basis}$, $N_\text{atom}$, and sparsity. Minimal basis sets are used to describe hydrogen, oxygen, and silicon atoms.}
\footnotesize
\begin{tabular}{c c c c c c c}
\hline
\hline
code & system & $\tilde{n}_\text{basis}$ & $N_\text{atom}$ & sparsity (\%) \\
\hline
DFTB+ & H$_2$O & 2 & 5,184 -- 41,472 & 87.6 -- 98.5 \\
DFTB+ &     Si & 4 & 2,000 -- 31,250 & 89.2 -- 99.2 \\
\hline
\hline
\end{tabular}
\label{tab:bulk}
\end{table*}

Calculations of the bulk set were performed on the Cray XC40 supercomputer Cori at NERSC. Architecture utilization is identical to those in Sec.~\ref{subsec:heavy}.

\begin{figure*}[ht!]
\centering
\includegraphics[width=0.66\textwidth]{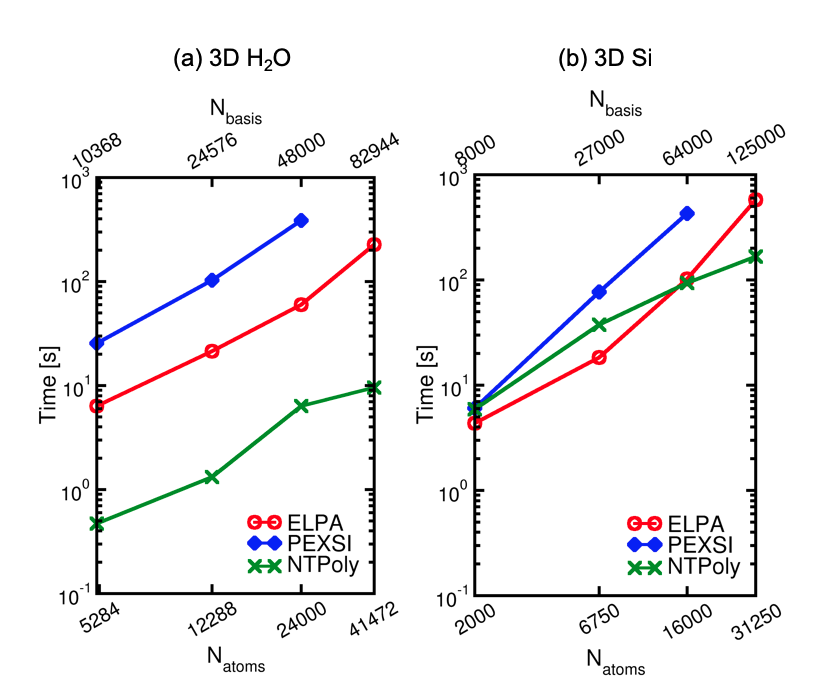}
\caption{Performance of key steps in ELPA, PEXSI, and NTPoly for (a) water models and (b) silicon models. ELPA computes all eigenvalues and eigenvectors. Results in this figure are obtained by running the DFTB+ code on the Cori supercomputer (Haswell partition) with 2,560 CPU cores (MPI tasks).}
\label{fig:bulk_d}
\end{figure*}

Fig.~\ref{fig:bulk_d} shows the performance of the ELPA, PEXSI, and NTPoly solvers in DFTB+ calculations of water (5,184 to 41,472 atoms; 10,368 to 82,944 basis functions) and silicon (2,000 to 31,250 atoms; 8,000 to 125,000 basis functions) models. Consistent with Figs.~\ref{fig:carbon_f}, \ref{fig:carbon_s}, and \ref{fig:heavy_f}, PEXSI is always slower than ELPA for these 3D structures. The NTPoly solver is able to outperform ELPA by an order of magnitude in all calculations of water molecules, as shown in Fig.~\ref{fig:bulk_d} (a). For the silicon case, NTPoly starts outperforming ELPA when the number of atoms becomes greater than 16,000 (64,000 basis functions). A speed-up of 4 can be achieved for 31,250 atoms (125,000 basis functions), and should be larger for more atoms.

It is worth noting that the density matrix purification parameters used in this subsection, i.e., matrix truncation threshold $10^{-5}$ and purification convergence criterion $10^{-2}$, are crucial to maintaining the sparsity of matrices in intermediate steps in Eq.~\ref{eq:ntpoly_purification}, and therefore achieving linear scaling computational cost. Density matrices computed from these parameters are sufficiently accurate for non-self-consistent calculations. The maximum difference in total energy is only 32.4 $\mu$eV/atom between results obtained with NTPoly and ELPA. Nevertheless, slower SCF convergence may be observed in self-consistent calculations, where tighter parameters need to be considered.

\begin{figure*}[ht!]
\centering
\includegraphics[width=0.66\textwidth]{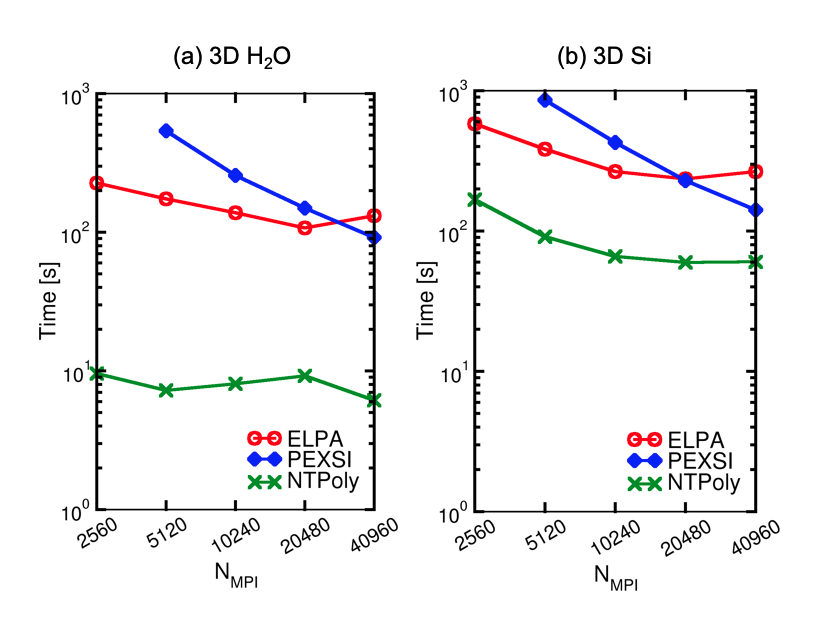}
\caption{Performance of key steps in ELPA, PEXSI, and NTPoly for (a) water 41,472-atom model and (b) silicon 31,250-atom model. ELPA computes all eigenvalues and eigenvectors. Results in this figure are obtained by running the DFTB+ code on the Cori supercomputer (Haswell partition).}
\label{fig:cpu_d}
\end{figure*}

We note that the PEXSI algorithm is highly scalable, as the poles in Eq.~\ref{eq:pexsi} can be evaluated fully in parallel. For each pole, the object $(\boldsymbol{H} - (z_l + \mu) \boldsymbol{S})^{-1}$ is computed by the PSelInv (parallel selected inversion) technique, which is able to make efficient use of thousands of CPU cores~\cite{pselinv_jacquelin_2016}. With only 20 CPU cores assigned to each pole, PEXSI calculations in Figs.~\ref{fig:carbon_f}, \ref{fig:carbon_s}, \ref{fig:heavy_f}, and \ref{fig:bulk_d} are far away from the scalability limit of the algorithm~\cite{pexsi_lin_2014,elsi_yu_2018}. To test the parallel performance of the solvers, we take the 41,472-atom water model and the 31,250-atom silicon model in Table~\ref{tab:bulk} and solve them using ELPA, PEXSI, and NTPoly with up to 40,960 CPU cores (MPI tasks). As shown in Fig.~\ref{fig:cpu_d}, the PEXSI solver exhibits a strong scaling superior to ELPA and NTPoly in both the water and silicon test cases. The ELPA solver is faster than PEXSI when using fewer than 20 thousand CPU cores, but it ceases to scale further and becomes slower than PEXSI when using more CPU cores. The NTPoly solver, although its strong scaling is not as good as PEXSI, is still the fastest solver for the two benchmark systems. The scalability of NTPoly may be extended by activating the 3D process grid feature for sparse matrix multiplications in NTPoly~\cite{ntpoly_dawson_2018}.

\section{Outlook}
\label{sec:outlook}
\subsection{Automatic Solver Selection}
\label{subsec:decision}
As an outcome of the benchmarks and analysis discussed in the previous section, we propose a semi-empirical mechanism to automatically select a solver for arbitrary problems. This mechanism is composed of two layers, a quick decision layer and a direct comparison layer. The former is intended to quickly estimate the solver performance using a few descriptors of a physical system, whereas the latter performs a more rigorous check in case that no quick decision can be made.

As Algorithm~\ref{alg:decision} shows, the quick decision layer takes four parameters of a physical system as its input, namely the system dimensionality ($N_\text{dim}$), the energy gap ($E_\text{gap}$), the number of basis functions employed ($N_\text{basis}$), and the matrix sparsity factor (defined as $N_\text{zero}/N_\text{basis}^2$ with $N_\text{zero}$ being the number of zero matrix elements). Three protocols are implemented in this layer: (1) ELPA is chosen for systems with fewer than 20,000 basis functions, regardless of the values of the other parameters. In benchmarks in Sec.~\ref{sec:results}, ELPA has demonstrated its efficiency for small-to-medium-sized systems. Although PEXSI or NTPoly could outperform ELPA for some system geometries and/or potentially different MPI library environments and processor counts even when $N_\text{basis} < 20,000$, the performance difference should be rather small and probably not critical. (2) When $N_\text{basis} > 20,000$, PEXSI is chosen for 1D/2D systems whose matrix sparsity factor is higher than 95\%. In our benchmarks, PEXSI consistently outperforms ELPA for large, sparse, low-dimensional systems thanks to its O($N$) and O($N^{1.5}$) scaling for 1D and 2D systems, respectively. Knowledge of the dimensionality of the system currently relies on user's input. (3) When $N_\text{basis} > 100,000$, NTPoly is chosen for systems with an energy gap larger than 0.5 eV and a sparsity factor higher than 99\%. The linear scaling density matrix purification methods in NTPoly rely on a proper energy gap, which ensures that the density matrix $\boldsymbol{P}$ is sufficiently sparse for large systems. The sparse matrix multiplication kernel in NTPoly efficiently utilizes this sparsity to maximize its performance. At present, the user is responsible for providing an estimate of the energy gap. When no estimate is available, the NTPoly solver will not be considered by the decision layer. Future plans include implementing fast, approximate algorithms to estimate the energy gap~\cite{dos_lin_2016,dos_dinapoli_2016}, and integrating more solvers into Algorithm~\ref{alg:decision}.

\begin{algorithm}[ht!]
\caption{Implementation of the quick decision layer for an automatic solver selection based on the analysis in Sec.~\ref{sec:results}. Input: system dimensionality $N_\text{dim}$ (1, 2, or 3), energy gap $E_\text{gap}$, number of basis functions $N_\text{basis}$, and matrix sparsity factor. Output: choice of solver.}
\begin{algorithmic}
\STATE{\textbf{function QUICK\_DECISION} ($N_\text{dim}$, $E_\text{gap}$, $N_\text{basis}$, sparsity, solver)}
\IF{($N_\text{basis} < 20,000$)}
    \STATE{solver = ELPA}
\ELSIF{($N_\text{dim} < 3$ \AND sparsity $> 95\%$)}
    \STATE{solver = PEXSI}
\ELSIF{($N_\text{basis} > 100,000$ \AND sparsity $> 99\%$ \AND $E_\text{gap} > 0.5$ eV)}
    \STATE{solver = NTPoly}
\ELSE
    \STATE{solver = NOT\_DECIDED}
\ENDIF
\RETURN{solver}
\end{algorithmic}
\label{alg:decision}
\end{algorithm}

The conditions in the three protocols above are checked sequentially. If the condition in the n$^\text{th}$ protocol has been satisfied, then a solver is chosen accordingly, and the (n+1)$^\text{th}$ and subsequent protocols will never be tested at all. There are cases where a quick decision cannot be made. For instance, a 3D, metallic system with more than 20,000 basis functions does not fall into any of the three protocols. In such cases, the direct comparison layer will take over. In this layer, we test candidate solvers one by one in order to figure out their performances relative to each other. Here, candidate solvers always include ELPA, may include PEXSI if the system is not 3D, and may include NTPoly if the system is highly sparse and not metallic. Suppose there are $N_\text{solver}$ candidate solvers, we iterate over them in the first $N_\text{solver}$ SCF steps, with different solvers being used in different SCF steps. Thus, after $N_\text{solver}$ SCF steps the timings for candidate solvers are measured, from which the optimal solver is identified. By incorporating the direct comparison layer into the SCF cycle, the overhead associated with the solver selection procedure is minimized.

\subsection{Reverse Communication Interface}
\label{subsec:rci}
Iterative eigensolvers are widely used in density-functional theory implementations based on planewave discretization, where the Hamiltonian matrix and the overlap matrix are applied as operators without being explicitly formed, as the sizes of these matrices are too large. Such an operator representation makes it less practical to design a single, universal interface of iterative eigensolvers for a wide range of planewave-based DFT codes, especially for those codes with OpenMP/MPI parallelization. Different DFT codes adopt different distribution patterns of wavefunctions. Implementing iterative solvers supporting a large variety of distribution patterns could still be considered, but such a design lacks extensibility. On the other hand, an interface that sticks to a particular distribution pattern would require conversions between different distribution patterns used by the interface and the user code, which might necessitate larger changes at the user code level and could also become a bottleneck in terms of time and/or memory. Therefore, iterative eigensolvers in ELSI are supported through a reverse communication interface (RCI) framework~\cite{rci_dongarra_1995} within the ELSI-RCI subproject -- i.e., as shown below, ELSI-RCI provides the algorithmic steps but allows a user code to retain control over matrix and vector distributions as well as other details of the linear algebra that is specific to the user code.

\begin{figure*}[ht!]
\centering
\includegraphics[width=0.33\textwidth]{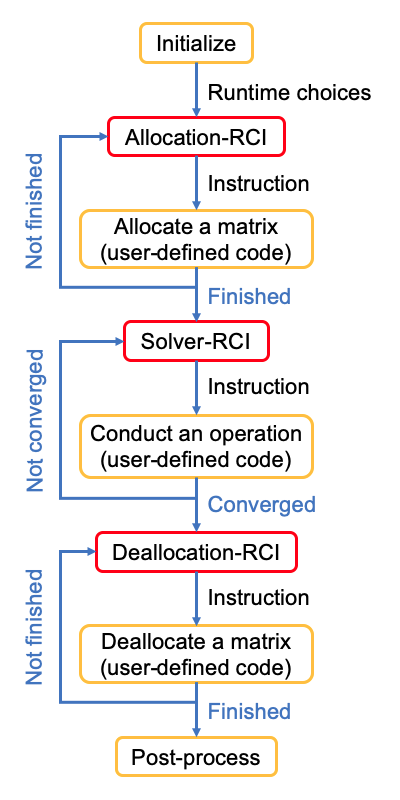}
\caption{Flow chart that describes the general workflow of ELSI-RCI. Yellow boxes: Zone of matrices and parallel operations implemented in user's driver. Red boxes: Zone of scalars and sequential operations implemented in ELSI-RCI. Black text outside boxes: Data structures being passed between functions. Depending on the user-specified runtime choices, ELSI-RCI gives instructions on the execution of an iterative eigensolver as well as the allocation/deallocation of matrices needed as workspace. The details of each RCI instruction must be defined and controlled by the user code.}
\label{fig:rci_flow}
\end{figure*}

One observation that is common to many iterative eigensolvers is that they only require a limited set of operations, mostly involving the application of the Hamiltonian and overlap operators and basic linear algebra operations such as matrix-vector multiplications. Different iterative eigensolvers only differ in the ordering of these operations (the only exception is the preconditioning step~\cite{preconditioner_knyazev_1998}). Since these operations are often already implemented in a DFT code, arranging them in a particular sequence actually yields an iterative eigensolver, and various solvers can be easily implemented by only altering the sequence of the operations. Following this idea, the current version of ELSI-RCI supports the Davidson method~\cite{davidson_davidson_1975,davidson_sleijpen_1996}, the orbital minimization method~\cite{libomm_corsetti_2014,omm_lu_2017}, the projected preconditioned conjugate gradient method~\cite{ppcg_vecharynski_2015}, and the Chebyshev filtering method~\cite{chebyshev_banerjee_2016,chebyshev_zhou_2006}. For the user-specified solver, ELSI-RCI provides a sequence of instructions, as shown in Fig.~\ref{fig:rci_flow}. ELSI-RCI has a data type \textbf{rci\_handle}, which is initialized by the user code. \textbf{rci\_handle} contains configurations of iterative eigensolvers, e.g., the solver type and the maximum number of iterations, and the status of an execution, e.g., the number of iterations so far. After initialization, ELSI-RCI conducts three stages of computations in a row: allocation stage, solver stage, and deallocation stage. Since different iterative eigensolvers use different numbers of temporary matrices of different sizes, the allocation stage and the deallocation stage are responsible for the allocation and deallocation of these temporary matrices. The allocation and deallocation instructions are given through \textbf{rci\_solve\_allocate} and \textbf{rci\_solve\_deallocate}, respectively. Each of these instructions should be implemented by the user code, i.e., the technical details such as matrix layouts etc. are intended to be defined and controlled by the user code.

The solver stage is the core of ELSI-RCI. It gives various instructions through \textbf{rci\_solve} with the order depending on the choice of eigensolver. Again, the technical details associated with each instruction should be implemented and thus controlled by the user code. Details of the instructions are encapsulated in a data type \textbf{rci\_instr}. Once an instruction is given, the user code executes it and returns nothing or a vector to ELSI-RCI. For most operations, no return from the user code is needed, while for some operations related to deflation, a vector of estimated eigenvalues is needed for ELSI-RCI to provide the next instruction. Such a two-step procedure is repeated until the prescribed stopping criterion is satisfied.

Instructions in the solver stage can be classified into three groups: controlling instructions, basic linear algebra instructions, and operator instructions. Controlling instructions include \textbf{RCI\_NULL}, \textbf{RCI\_CONVERGE}, and \textbf{RCI\_STOP}, which indicate that no operation is needed, the chosen iterative eigensolver has converged, and the chosen solver has stopped without reaching convergence, respectively. Basic linear algebra instructions include some of the basic BLAS-level and LAPACK-level operations. Operator instructions include \textbf{RCI\_H\_MULTI} for the application of the Hamiltonian operator, \textbf{RCI\_S\_MULTI} for the application of the overlap operator, and \textbf{RCI\_P\_MULTI} for the application of the preconditioner. Since different preconditioners are preferred for different iterative eigensolvers, the user code is currently expected to prepare and provide preconditioners for the chosen solver to achieve optimal performance.

On the user side, a driver for ELSI-RCI is required, which performs allocation, deallocation, basic linear algebra operations, and operator applications per the instruction from ELSI-RCI. Since almost all these operations are standard parts of existing DFT codes, constructing an ELSI-RCI driver using existing code should be rather straightforward. With this driver, users are then able to call different iterative eigensolvers in ELSI-RCI with a simple change of the solver name in the initialization step of ELSI-RCI. This provides a unified access to a variety of iterative solvers through a single interface.

\subsection{Towards GPU-Accelerated High-Performance Computing}
\label{subsec:gpu}
The past ten years witnessed significant growth in the usage of GPU accelerators in high-performance computing (HPC). According to the TOP500 list~\cite{top500}, the number of GPU-accelerated supercomputers skyrocketed from two in 2010 to over a hundred in the most recent release of the list (November 2019). Representatives of state-of-the-art GPU supercomputers include Summit at Oak Ridge National Laboratory~\cite{summit} and Sierra at Lawrence Livermore National Laboratory~\cite{sierra}. Combining IBM POWER9 CPUs and NVIDIA Volta GPUs, Summit and Sierra rank 1$^\text{st}$ and 2$^\text{nd}$, respectively, on the current TOP500 list. They, as well as many other supercomputers with GPU accelerators, allow for a substantial boost in performance and power efficiency compared to traditional CPU-only machines.

The computation power from GPUs has been utilized by the electronic structure community for larger and faster simulations~\cite{gpu_yasuda_2008a,gpu_yasuda_2008b,gpu_ufimtsev_2008,gpu_ufimtsev_2009a,gpu_ufimtsev_2009b,gpu_genovese_2009,gpu_maintz_2011,gpu_hacene_2012,gpu_titov_2013,gpu_jia_2013a,gpu_jia_2013b,gpu_ratcliff_2018,gpu_huhn_2020}. Various eigensolver and density matrix solver implementations targeting hybrid CPU-GPU machines have been reported. The GPU-enabled ELPA one-stage solver outperforms the CPU version on a few compute nodes when the matrix size exceeds several thousands~\cite{elpa_kus_2019}. Other publicly available GPU-enabled eigensolvers~\cite{magma_tomov_2010,magma_dongarra_2014,cusolver}, to our knowledge, are currently limited to shared memory executions on a single compute node. The 2$^\text{nd}$ order trace resetting density matrix purification method was ported to GPUs~\cite{purification_cawkwell_2012} by using matrix multiplications provided in the cuBLAS library~\cite{cublas}. Similarly, the GPU-accelerated sparse matrix linear algebra routines in the DBCSR library~\cite{dbcsr_borstnik_2014,dbcsr_lazzaro_2017} can be employed to compute the density matrix on GPUs.

Several challenges are involved in developing full-fledged eigensolvers and density matrix solvers running on distributed memory, massively parallel, GPU-accelerated supercomputers. MPI communications involving GPUs are more expensive than those between CPUs. GPUs implement a deep memory hierarchy which requires data to be copied out from GPU to CPU to participate in MPI communications. Therefore, existing algorithms must be redesigned to reduce CPU-GPU and GPU-GPU data communications, and to overlap communications with computations as much as possible. In the context of dense eigensolvers, GPUs may perform large amounts of matrix-matrix operations, especially in the two-stage tridiagonalization~\cite{elpa_marek_2014,2step_bischof_1994}, to amortize the cost of data transfers. However, the back-transformation of eigenvectors in the two-stage algorithm is not naturally suited for GPU acceleration, as it lacks an explicit data-parallel computation pattern to be executed on a large amount of GPU cores. On the other hand, several sparse matrix computation techniques do not display large contiguous blocks of data on which GPUs can efficiently operate. Sparse matrix algebra relies heavily on indirect addressing to access data in the sparse matrix. This type of memory access does not generally perform well on GPUs, which prefer memory accesses to be organized in a very specific way, called coalescing, for maximum performance.

The situation is changing with the arrival of new GPU supercomputers such as Summit and Sierra. The new generation of NVIDIA Volta GPUs provides unprecedented memory bandwidth and data transfer speed compared to its predecessors. Intra-node data movements can take advantage of the NVLink technology~\cite{nvlink_foley_2017} for high-bandwidth interconnect between GPUs and between GPUs and CPUs. Inter-node data movements can benefit from CUDA-aware MPI, which is now supported by the hardware. In addition, the large cache of Volta GPUs has the potential to substantially accelerate indirect addressing in sparse matrix computations. All of these features make platforms like Summit and Sierra good candidates for accelerating eigensolvers and density matrix solvers. We are planning to leverage them within ELSI. For eigenproblems on individual nodes, the MAGMA library~\cite{magma_tomov_2010,magma_dongarra_2014} has long constituted the state of the art. Additionally, we plan to port and optimize the distributed-parallel ELPA two-stage eigensolver (ELPA2) and the PEXSI and NTPoly density matrix solvers to GPU platforms. In fact, a separate benchmark paper for GPU acceleration in the ELPA one-stage eigensolver already exists~\cite{elpa_kus_2019}. Development of GPU-accelerated ELPA2~\cite{elpa_yu_2020} and PEXSI is ongoing, and will be reported separately by some of the authors.

\section{Conclusions}
\label{sec:conclude}
In this paper, we summarize recent developments of the ELSI electronic solver interface. Among all the upgraded and new features, the following are highlighted:

\begin{enumerate}
\item New solvers, namely the upgraded PEXSI density matrix solver, the linear scaling density matrix solver in the NTPoly library, the parallel spectrum slicing sparse eigensolver in the SLEPc library, the penta-diagonalization-based dense eigensolver in the EigenExa library, and the GPU-accelerated, shared-memory dense eigensolvers in the MAGMA library.

\item Two new matrix formats. The SIESTA\_CSC format has greatly simplified the integration of ELSI into the SIESTA and DFTB+ electronic structure code packages. The GENERIC\_COO format offers maximal flexibility of matrix distribution. It is expected to aid in the integration of ELSI with packages using a custom matrix storage layout.

\item A backward-compatible extension of the interface that allows for parallel calculations of spin-polarized and/or periodic systems.

\item Several routines serving geometry optimization and molecular dynamics calculations. This includes calculation of the energy-weighted density matrix, extrapolation of the density matrix and wavefunctions, and a ``smart'' reinitialization of ELSI which reuses information across geometry steps.

\item Efficient parallel matrix I/O routines based on MPI I/O.

\item Standardized JSON output via the FortJSON library.
\end{enumerate}

Furthermore, we have assessed the performance of three electronic structure solvers, ELPA, PEXSI, and NTPoly, by running a systematic set of benchmark calculations with both Kohn--Sham density-functional theory and density-functional tight-binding theory. Unsurprisingly, the performance of the solvers depends on the specifics of the problem. For small-to-medium-sized structures up to several hundreds of atoms, the highly optimized dense eigensolver ELPA is always a stable and efficient solution. As the system size increases, the Hamiltonian, overlap, and density matrices become more sparse, rendering lower scaling methods based on sparse linear algebra more favorable. In particular, the pole expansion and selected inversion method PEXSI is best-suited for low dimensional systems. It also exhibits a nearly ideal parallel scalability for at least 40 thousand CPU cores. For large systems with an energy gap, a speed-up over ELPA can be achieved by using the density matrix purification algorithms in NTPoly. These results clearly identify the regimes where sparse density matrix solvers can beat diagonalization, based on which we propose a semi-empirical mechanism to automate the selection of the optimal solver for a given problem. The O($N^3$) diagonalization bottleneck in large-scale electronic structure simulations can thus be alleviated to some extent.

On the other hand, our results imply that reaching the crossover point between the diagonalization method and the state-of-the-art density matrix methods would still require many hundreds of, or even thousands of atoms. Nevertheless, the ELSI interface offers a platform for developing, validating, and comparing algorithms, which has been a driving force of improvements in eigensolvers and density matrix solvers. We expect this collaborative effort to continue to lower the computational cost of large-scale electronic structure theory. Our ongoing construction of a reverse communication interface (RCI) framework for iterative eigensolvers will hopefully facilitate the optimal use of eigensolvers in planewave-based density-functional theory implementations. We also hope to support dedicated optimizations targeting GPU architectures, which now seem critical for the realization of exascale electronic structure calculations in the future.

\section*{Acknowledgments}
\label{sec:thanks}
This research was supported by the National Science Foundation (NSF) under Award No. 1450280. This research used resources of the National Energy Research Scientific Computing Center (NERSC), a U.S. Department of Energy (DOE) Office of Science User Facility operated under Contract No. DE-AC02-05CH11231. This research also used resources of the Argonne Leadership Computing Facility, a DOE Office of Science User Facility supported under Contract No. DE-AC02-06CH11357. We appreciate the constructive feedback from many fellow researchers in the electronic structure community, including developers and users of the BigDFT, CP2K, DFTB+, DGDFT, FHI-aims, NTChem, SIESTA projects and of CECAM's Electronic Structure Library project. Yu was additionally supported by a fellowship from the Molecular Sciences Software Institute under NSF Award No. 1547580. Garc\'{i}a thanks EU H2020 grant 824143 (``MaX: Materials at the eXascale'' CoE), Spain's AEI (PGC2018-096955-B-C44 and ``Severo Ochoa'' grant SEV-2015-0496), and GenCat 2017SGR1506.

\appendix
\section{ELSI Build System}
\label{app:cmake}
Building of the ELSI library is managed by the CMake build system generator. CMake is a cross-platform free and open-source software with support for several build systems (e.g., GNU Make). The build process consists of largely two steps: the configuration step, which generates the build files (e.g., makefiles for GNU Make), and the compilation step.

The requirements for building ELSI are (as of its 2.5.0 release):
\begin{itemize}
\item CMake version $\ge$ 3.0
\item Fortran compiler (Fortran 2003 compliant)
\item C compiler (C99 compliant)
\item MPI
\item Linear algebra libraries (BLAS, LAPACK, ScaLAPACK)
\end{itemize}
Additionally, some optional requirements are:
\begin{itemize}
\item C++ compiler (C++11 compliant, for PEXSI support)
\item SLEPc version 3.13 (for SLEPc-SIPs support)
\item PETSc version 3.13 (for SLEPc-SIPs support)
\item EigenExa version 2.4 (for EigenExa support)
\item MAGMA version 2.5 (for MAGMA support)
\item BSEPACK version 0.1 (for BSEPACK support)
\end{itemize}

During configuration, user-configurable settings for the project are stored in the CMake cache. The cache variables are persistent between builds, allowing the user to make changes to the build configuration with minimal recompilation of the source files. One method of setting the cache variables is to store them in a dedicated file. The command for configuring ELSI would then look as follows:
\begin{verbatim}
cmake -C <initial_cache.cmake> <elsi>
\end{verbatim}
where \verb+<initial_cache.cmake>+ is the location of the initial cache file and \verb+<elsi>+ is the location of the ELSI source root directory. A template for an ELSI initial cache file is given below, showing the common variables that the user needs to set. Examples for different combinations of compilers and external libraries are shipped with the ELSI package. The full list of cache variables applicable to ELSI are detailed in the ELSI documentation. In order to ensure that the user is aware of all the performance-critical build details, compilers and external libraries are not automatically detected. If the user wants to link against, e.g., linear algebra libraries, then these need to be explicitly specified among the cache variables.
\begin{footnotesize}
\begin{verbatim}
# An example CMake initial cache file for ELSI
set(CMAKE_Fortran_COMPILER "mpifort" CACHE STRING "MPI Fortran compiler")
set(CMAKE_C_COMPILER "mpicc" CACHE STRING "MPI C compiler")
set(CMAKE_CXX_COMPILER "mpicxx" CACHE STRING "MPI C++ compiler")
set(CMAKE_Fortran_FLAGS "-O3" CACHE STRING "Fortran flags")
set(CMAKE_C_FLAGS "-O3" CACHE STRING "C flags")
set(CMAKE_CXX_FLAGS "-O3" CACHE STRING "C++ flags")
set(ENABLE_PEXSI ON CACHE BOOL "Enable PEXSI")
set(ENABLE_TESTS ON CACHE BOOL "Enable tests")
set(LIB_PATHS "/path/to/external/libraries" CACHE STRING "Library paths")
set(LIBS "scalapack lapack blas" CACHE STRING "External libraries")
\end{verbatim}
\end{footnotesize}

By default, all the solver libraries that are redistributed with ELSI are built. There is also the possibility of linking ELSI against any of those libraries compiled externally. This can be beneficial, e.g., for testing the latest upstream changes that are not yet present in a given version of ELSI. As an example, for using ELPA externally, one could turn on \verb+USE_EXTERNAL_ELPA+ in the initial cache file and adjust the \verb+INC_PATHS+, \verb+LIB_PATHS+, and \verb+LIBS+ variables for including ELPA.

After the configuration step, ELSI may be built by
\begin{verbatim}
make
\end{verbatim}
when using GNU Make, or a more generic command
\begin{verbatim}
cmake --build <build>
\end{verbatim}
where \verb+<build>+ is the build directory.

One of the major benefits of using CMake is that ELSI can be easily included in other CMake projects. Only two lines are necessary to link against ELSI,
\begin{verbatim}
find_package(elsi <version> REQUIRED PATHS <elsi_install>)
target_link_libraries(my_project PRIVATE elsi::elsi)
\end{verbatim}
All the necessary libraries and directories with header and modules files, including the main ELSI library, propagate with the single \verb+elsi::elsi+ target. The optional argument \verb+<version>+ specifies the minimum required version of ELSI. This helps to avoid potential API conflicts with earlier ELSI versions. The arguments \verb+REQUIRED+ and \verb+PATHS+ specify whether to stop processing if the package is not found and where to search for the ELSI installation, respectively. If ELSI is installed in a standard system location, the \verb+PATHS+ argument may be omitted. In order to ensure that ELSI is found only at the specified location, the \verb+NO_DEFAULT_PATH+ option of \verb+find_package+ may be used.

\bibliographystyle{elsarticle-num}
\bibliography{elsi2_bib}

\end{document}